\documentclass[aps,preprint,showpacs,preprintnumbers,amsmath,amssymb]{revtex4}
\usepackage[dvips]{graphicx}
\usepackage{float}
\begin{document}

\title{Fermion production in a magnetic field in a de Sitter Universe}
\author{Cosmin Crucean \thanks{E-mail:~~crucean@physics.uvt.ro}
and Mihaela-Andreea B\u aloi \thanks{E-mail:~~mihaela.baloi88@e-uvt.ro}\\
{\small \it West University of Timi\c soara,}\\
{\small \it V. Parvan Avenue 4 RO-300223 Timi\c soara,  Romania}}

\begin{abstract}
The process of fermion production in the field of a magnetic dipole on a de Sitter expanding universe is analyzed. The amplitude and probability for production of massive fermions are obtained using the exact solution of the Dirac equation written in the momentum-helicity basis. We found that the most probable transitions are those that generate the fermion pair perpendicular to the direction of the magnetic field. The behavior of the probability is graphically studied for large/small values of the expansion factor, and a detailed analysis of the probability in terms of the angle between the momenta vectors of the particle and antiparticle is performed. The phenomenon of fermion production is significant only at large expansion which corresponds to the conditions from the early Universe. When the expansion factor vanishes we recover the Minkowski limit where this process is forbidden by the simultaneous energy-momentum conservation.
\end{abstract}


\maketitle

\section{Introduction}
The problem of particle production in electric fields in a de Sitter background is discussed in the literature using both perturbative and nonperturbative methods \cite{23,31,32}. These studies take into account the production of scalar particles and Dirac particles in external electric fields. The main results of these papers concern the calculation of the density number of produced particles \cite{23,31,32}. Another important conclusion that arises from these papers is that the particles are emitted parallel with the direction of the electric field \cite{23,31,32}. Comparatively the problem of pair production in magnetic fields in de Sitter geometry received less attention. In fact, a direct perturbative/nonperturbative calculation of fermion pair production in a magnetic field in de Sitter geometry was not performed in the literature to the best of our knowledge. However, in Minkowski space the production of pairs in strong magnetic fields was discussed in a few papers, using nonperturbative methods \cite{8,9}. In Ref. \cite{8}, the production of neutral fermions in inhomogeneous magnetic fields was studied. It is known that such strong magnetic fields could exist in the Universe. But the neutron stars or black holes that are supposed to have such magnetic fields also have a strong gravitational field, which could generate particle production. This is the reason we consider that it is important to understand how the calculations from flat space will be affected by the presence of a gravitational field. The problem of the origin of magnetic fields in the Universe was studied in a series of papers \cite{2,3,4,5,6,7,9,11}. This subject is still open since for a complete explanation and understanding of this problem, one needs to combine hydrodynamics with the theory of general relativity. When a gravitational field is present, the problem of pair production in a magnetic field becomes more complicated since there are technical problems that need to be considered with care. For example, if the Dirac field is coupled with an external potential that generates a magnetic field, the resulting Dirac equation is not easy to be solved if the geometry is not flat. In this paper we use perturbative methods to study fermion production by a magnetic field when a gravitational field is present. More precisely we consider the field of a magnetic dipole placed in the large expansion conditions of the early Universe. The approach in other geometries such as Schwarzschild seems to be more difficult because there are no exact solutions of the field equations. For this reason one needs to be careful when trying to extend some of the conclusions reached in this paper to the astronomical objects that have a magnetic field and strong gravitational fields that in principle could generate particle production even for the present day expansion of the space. Therefore we must specify that our results are valid for strong gravitational fields of the early Universe and weak dipole magnetic fields and that for the present day expansion of the space the probability obtained in this paper is vanishing.

The problem of pair production in de Sitter geometry was studied using nonperturbative methods \cite{10,14,24,25,26,27,31,32,35,36} as well as perturbative methods \cite{mo,20,21,28,29,30,33,34}. In flat space a direct perturbative calculation of pair production in an external field have the amplitude of transition zero since the energy-momentum conservation forbids this process \cite{15,16,17}. In de Sitter geometry this is no longer valid, and we expect a direct perturbative calculation to give nonvanishing amplitudes of transition for the first order processes that generate production in external electromagnetic fields. In this paper we will use the recently developed QED formalism on de Sitter spacetime \cite{20,21} for computing the first order amplitude that corresponds to the process $vacuum\rightarrow e^++e^-$, in the presence of the field of a magnetic dipole. The time reversed process represents the absorption of the fermion pair by the magnetic field. This study is performed using the exact solutions of the Dirac equation in the momentum-helicity basis \cite{19,22,37}. Adopting the QED formalism in de Sitter geometry \cite{21}, one can study the behavior of the transition amplitudes/probabilities at large expansion (early Universe), as well as the Minkowski limit where the expansion factor vanishes. It is also possible, using this approach, to study what happens in the interesting case when gravity becomes weak.

The paper is organized as follows. In the second section we present the basic elements about the solution of the Dirac equation in de Sitter geometry and give the expression of the potential that produces the field of a magnetic dipole. In the third section we remind the reader the definition of the transition amplitude in the first order of perturbation theory for de Sitter QED. Then we present the main steps for computing the transition amplitude/probability, and we discuss the physical consequences of our calculations. In the fourth section the behavior of the probabilities is graphically studied in terms of the expansion parameter. Also in this section we analyze the probability in terms of the angle between momenta vectors of the pair, for a given expansion parameter. Section five is dedicated to the limit cases as the expansion parameter varies from large to small values, relative to the particle mass. In section six we present our conclusions, and in the Appendix we give the main steps that help us to compute the amplitude and probability.

\section{Free fields}
We will start in this section with a brief review of the exact solutions
of the free Dirac and Maxwell equations in de Sitter geometry, expanded in the momentum-helicity basis.
The line element \cite{0} for the de Sitter universe is:
\begin{equation}\label{metr}
ds^{2}=dt^{2}-e^{2\omega t}d\vec{x}\,^{2}=\frac{1}{(\omega t_{c})^2}(dt_{c}^{2}-d\vec{x}\,^{2})
\end{equation}
where $\omega$ is the expansion factor and $\omega>0$, while the conformal time $t_{c}=-\frac{1}{\omega}\,e^{-\omega t}$.  To define half-integer spin fields on curved spacetime it is required to use the tetrad fields \cite{2} $e_{\widehat{\mu}}(x)$ and $\widehat{e}^{\widehat{\mu}}(x)$,
which fix the local frames and corresponding coframes and which are labeled by the local
indices $\widehat{\mu},\widehat{\nu},...=0,1,2,3$. For the
line element (\ref{metr}), we can choose the Cartesian gauge with
the nonvanishing tetrad components,
\begin{equation}
e^{0}_{\widehat{0}}=e^{-\omega t}  ;\,\,\,e^{i}_{\widehat{j}}=\delta^{i}_{\widehat{j}}\,e^{-\omega t}.
\end{equation}

The Dirac equation in de Sitter geometry was studied in \cite{19,36,37}. In \cite{19} the solutions of the massive Dirac field were obtained in the momentum-helicity basis.
The positive/negative frequency modes of momentum $\vec{p}$ and
helicity $\lambda$, derived in \cite{19}, are:
\begin{eqnarray}\label{sol}
&&U_{\vec{p},\lambda}(t,\vec{x}\,)=\frac{\sqrt{\pi
p/\omega}}{(2\pi)^{3/2}}\left (\begin{array}{c}
\frac{1}{2}e^{\pi k/2}H^{(1)}_{\nu_{-}}(\frac{p}{\omega} e^{-\omega t})\xi_{\lambda}(\vec{p}\,)\\
\lambda e^{-\pi k/2}H^{(1)}_{\nu_{+}}(\frac{p}{\omega} e^{-\omega
t})\xi_{\lambda}(\vec{p}\,)
\end{array}\right)e^{i\vec{p}\cdot\vec{x}-2\omega t};\nonumber\\
&&V_{\vec{p},\lambda}(t,\vec{x}\,)=\frac{\sqrt{\pi
p/\omega}}{(2\pi)^{3/2}} \left(
\begin{array}{c}
-\lambda\,e^{-\pi k/2}H^{(2)}_{\nu_{-}}(\frac{p}{\omega} e^{-\omega t})\,
\eta_{\lambda}(\vec{p}\,)\\
\frac{1}{2}\,e^{\pi k/2}H^{(2)}_{\nu_{+}}(\frac{p}{\omega} e^{-\omega t}) \,\eta_{\lambda}(\vec{p}\,)
\end{array}\right)
e^{-i\vec{p}\cdot\vec{x}-2\omega t},
\end{eqnarray}
where $H^{(1)}_{\nu}(z), H^{(2)}_{\nu}(z)$ are Hankel functions of the first and second kinds and $k=\frac{m}{\omega},\nu_{\pm}=\frac{1}{2}\pm ik$. The unit normalized helicity spinors are given by:
\begin{equation}\label{xi}
\xi_{\frac{1}{2}}(\vec{p}\,)=\sqrt{\frac{p_3+p}{2 p}}\left(
\begin{array}{c}
1\\
\frac{p_1+ip_2}{p_3+p}
\end{array} \right)\,,\quad
\xi_{-\frac{1}{2}}(\vec{p}\,)=\sqrt{\frac{p_3+p}{2 p}}\left(
\begin{array}{c}
\frac{-p_1+ip_2}{p_3+p}\\
1
\end{array} \right)\,,
\end{equation}
while $\eta_{\sigma}(\vec{p}\,)= i\sigma_2
[\xi_{\sigma}(\vec{p}\,)]^{*}$. These spinors satisfy the relation:
\begin{equation}\label{pa}
\vec{\sigma}\vec{p}\,\xi_{\lambda}(\vec{p}\,)=2p\lambda\xi_{\lambda}(\vec{p}\,)
\end{equation}
with $\lambda=\pm1/2$, where $\vec{\sigma}$ are the Pauli
matrices and $p=\mid\vec{p}\mid$ is the modulus of the momentum vector. Since the charge conjugation symmetry remains valid in a curved background, the negative frequency modes are \cite{19}:
\begin{equation}\label{charge}
V_{\vec{p},\lambda}(x)=i\gamma^{2}\gamma^{0}(\bar{U}_{\vec{p},\lambda}(x))^{T}.
\end{equation}

These solutions are normalized such that \cite{19}:
\begin{equation}\label{ot}
\int d^{3}x
(-g)^{1/2}\bar{U}_{\vec{p},\lambda}(x)\gamma^{0}U_{\vec{p}\,\,',\lambda\,'}(x)=
\delta_{\lambda\lambda\,'}\delta^{3}(\vec{p}-\vec{p}\,').
\end{equation}
Also the normalization relation (\ref{ot}) is fulfilled by the $V$ solutions, with the specification that the two modes are orthogonal.
To clarify what mean positive/negative frequencies we can use the asymptotic expansion of the Hankel functions at large $z$ \cite{12,13,18},
\begin{equation}
H^{(1,\,2)}_{\nu}(z)\simeq\left(\frac{2}{\pi z}\right)^{1/2}e^{\pm i(z-\nu\pi/2-\pi/4)},
\end{equation}
and obtain that for $t\rightarrow-\infty$, the modes defined above behave as a positive/negative frequency modes with respect to conformal time $t_{c}=-e^{-\omega t}/\omega$:
\begin{equation}
U_{\vec{p},\lambda}(t,\vec{x}\,)\sim e^{-ipt_{c}};\,\,V_{\vec{p},\lambda}(t,\vec{x}\,)\sim e^{ipt_{c}}.
\end{equation}\label{bd}
This is precisely the behavior in the infinite past for positive/negative frequency modes, which defines the Bunch-Davies vacuum \cite{12}.
\par

Further we establish the equation for the external field in de Sitter geometry. In classical electrodynamics the expression of the field of a magnetic dipole \cite{1}
\begin{equation}\label{cm}
\vec{B}_{M}=\nabla\times\vec{A}_{M}=\frac{\vec{x}\,(\vec{\mathcal{M}}\cdot\vec{x}\,)-\vec{\mathcal{M}}(\vec{x}\cdot\vec{x}\,)}{|\vec{x}|^5}
\end{equation}
is obtained from the vector potential \cite{1}:
\begin{equation}\label{am}
\vec{A}_{M}=\frac{\vec{\mathcal{M}}\times\vec{x}}{|\vec{x}|^3},
\end{equation}
where $\vec{\mathcal{M}}$
is the magnetic dipole moment. Now recalling the conformal invariance of Maxwell equations, it is simple to establish the expression of $\vec{A}$ in de Sitter geometry. If $\vec{A}_{M}$ is the vector potential in Minkowski space, then at a conformal transformation the vector potential in de Sitter geometry can be expressed as follows \cite{22}:
\begin{equation}
A^{\mu}=\Omega^{-1}A_{M}^{\mu},
\end{equation}
where $\Omega=(\omega t_{c})^{-2}$ for the metric given in Eq.(\ref{metr}). In a local frame the expression of the vector potential is obtained using the simple relation $A^{\widehat{\mu}}=e^{\widehat{\mu}}_{\nu}A^{\nu}$.
Then taking $A^{0}=0$ and using $A^{\widehat{i}}=e^{\widehat{i}}_{j}A^{j}$ one finds that in a local frame the de Sitter vector potential that generates the magnetic field has the following expression:
\begin{equation}\label{pot}
\widehat{\vec{A}}(x)=\frac{\vec{\mathcal{M}}\times\vec{x}}{|\vec{x}|^3}\,  e^{-\omega
t},A^{\widehat{0}}(x)=0.
\end{equation}
The hatted indices indicate the components of the four vector potential
in local frames.

\section{Probability of fermion production}
The calculation of the transition amplitude is based on the QED formalism in de Sitter geometry constructed in \cite{21}. This formalism uses perturbative methods to establish the form of the transition amplitudes as in the Minkowski QED. For pair production in an external electromagnetic field, supposing that the fields are coupled by the elementary electric charge $e$, the expression of the transition amplitude reads \cite{20,21}:
\begin{equation}\label{ampl}
\mathcal{A}_{e^-e^+}=-ie \int d^{4}x
\left[-g(x)\right]^{1/2}\bar U_{\vec{p},\,\lambda}(x)\gamma_{\mu}A^{\widehat{\mu}}(x) V_{\vec{p}\,\,',\,\lambda'}(x)
\end{equation}
where $e$ is the unit charge of the field and $U_{\vec{p},\,\lambda}(x)\,\,,V_{\vec{p}\,\,',\,\lambda'}(x)$ are the exact solutions of the Dirac equation in the momentum-helicity basis given in the previous section. By setting $A^{\widehat{0}}(x)=0$, the only remaining term that gives a contribution to the amplitude (\ref{ampl}) contains the spatial part of the vector potential,
\begin{equation}\label{ampl1}
\mathcal{A}_{e^-e^+}=-ie \int d^{4}x
\left[-g(x)\right]^{1/2}\bar U_{\vec{p},\,\lambda}(x)\vec{\gamma}\cdot\widehat{\vec{A}}(x) V_{\vec{p}\,\,',\,\lambda'}(x).
\end{equation}
Related to the above QED perturbative approach to the problem of particle production in de Sitter space, there are a few observations that need to be made. First, we must point out that we work in the chart with conformal time $t_{c}\in(-\infty,0)$, which covers only half of the whole de Sitter manifold. This chart covers the expanding portion of de Sitter space, and the perturbative QED presented in \cite{21} is also constructed in this chart. We must specify that the phenomenon of particle production due to the interaction was first studied in \cite{14,fo}. A detailed presentation for this formalism can be found in \cite{14}, and we must stress that our amplitude (\ref{ampl}) is equivalent with the in-in amplitude \cite{14,fo}. That means that we take the in vacuum to be the same as the out vacuum, and when we speak about the notion of particle, we refer to the in modes. The first examples of processes that take into account the influence of the field interaction upon the particle production in an expanding background can be found in \cite{fo}. In \cite{14,fo}, it was shown that the phenomenon of particle production due to the fields interaction could become more important in comparison with the pure gravitational field production. For these reasons we believe that for a complete picture related to the phenomenon of particle production in expanding de Sitter space, one must complete the existing cosmological production with the production that results from fields interactions. The last important observation is that the outcome of our calculations will be the transition probability that can be interpreted in terms of the density number of particles as in \cite{fo}. But there is still plenty of work to be done if one wants to obtain measurable quantities or to set up one experiment for measuring the probability. For example, in a scattering process, if one wants to define the cross section in de Sitter space, there are many difficulties that appear. These are related to the fact that the incident flux will be a quantity dependent on time, and in addition, as a result of the loss of the energy-momentum conservation, the usual delta Dirac term will be missing from the amplitude/probability. As a result, the cross section will depend on the form of the incident wave, which seems to be an unusual situation. All these problems need to be addressed for completing the perturbative results obtained so far for the quantum field theory in de Sitter space.

There are also important results related to the problem of infrared and ultraviolet divergences that could appear in quantum field theory on de Sitter space \cite{mo1,mo2,ak}. These studies deal with the divergences that appear in the two point function in the case of a scalar field. An interesting result related to the quantum field theory on de Sitter space was obtained in \cite{mo}, where a S-matrix formalism was constructed for weakly coupled quantum field theories in the global de Sitter manifold. The main results of \cite{mo} prove that the S-matrix is unitary, is de Sitter invariant and transforms properly under CPT. Moreover, the S-matrix obtained in \cite{mo} reduces to the usual S-matrix from Minkowski space in the flat-space limit. Here we restrict study to the influence of the fields interaction on the particle production in the chart that covers only the expanding portion of de Sitter space.

\subsection{The amplitude}
Using the spinor solutions (\ref{sol}) and the vector potential given in (\ref{pot}) the amplitude could be written in terms of a spatial integral and a temporal integral. The result of the spatial integral reads:
\begin{equation}
\int d^3x\frac{\vec{x}}{|\vec{x}|^3}\,e^{-i(\vec{p}+\vec{p}\,')\vec{x}}=-\frac{4\pi i(\vec{p}+\vec{p}\,')}{|\vec{p}+\vec{p}\,'|^2}.
\end{equation}
The temporal integral can be expressed using the relation between Hankel functions and Bessel $K$ functions \cite{12,13} and changing the integration variable to $z=e^{-\omega t}/\omega$. Collecting all results, the transition amplitude can be brought to the form:
\begin{eqnarray}
\mathcal{A}_{e^-e^+}&=&\frac{ie\sqrt{pp\,'}}{2\pi^{3}|\vec{p}+\vec{p}\,'|^{2}}\left[-sgn(\lambda\lambda\,')\int_{0}^{\infty}dz z K_{\nu_{-}}(ipz)K_{\nu_{-}}(ip\,'z)
+\int_{0}^{\infty}dz z K_{\nu_{+}}(ipz)K_{\nu_{+}}(ip\,'z)\right]\nonumber\\
&&\times\xi_{\lambda}^{+}(\vec{p}\,)[\vec{\sigma}\cdot(\vec{\mathcal{M}}\times(\vec{p}+\vec{p}\,\,'))]\eta_{\lambda\,'}(\vec{p}\,\,'),
\end{eqnarray}
where $sgn$ stands for the sign function.
These integrals can be solved in terms of Heaviside step function $\theta$, gamma Euler functions $\Gamma$, and hypergeometric functions $_{2}F_{1}$, as we show in the Appendix.
The final form of the transition amplitude for fermion pair production in the field of a magnetic dipole in de Sitter space reads:
\begin{eqnarray}\label{af}
\mathcal{A}_{e^-e^+} &=& -\frac{ie}{4\pi^{3}|\vec{p}+\vec{p}\,'|^{2}}\biggl\{ \frac{\theta(p-p\,')}{p}\left[f_{k}^{*}\left(\frac{p\,'}{p}\right)-sgn(\lambda\lambda\,')f_{k}\left(\frac{p\,'}{p}\right)
\right]\nonumber\\
&&+\frac{\theta(p\,'-p)}{p\,'}\left[f_{k}^{*}\left(\frac{p}{p\,'}\right)-sgn(\lambda\lambda\,')f_{k}\left(\frac{p}{p\,'}\right)\right]\biggl\}
\xi_{\lambda}^{+}(\vec{p}\,)[\vec{\sigma}\cdot(\vec{\mathcal{M}}\times(\vec{p}+\vec{p}\,'))]\eta_{\lambda\,'}(\vec{p}\,\,'),
\nonumber\\
\end{eqnarray}
where the functions $f_{k}\left(\frac{p\,'}{p}\right)$ are defined as follows:
\begin{eqnarray}\label{ff}
f_{k}\left(\frac{p\,'}{p}\right)&=&\left(\frac{p\,'}{p}\right)^{1-ik}\Gamma\left(\frac{3}{2}-ik\right)\Gamma\left(\frac{1}{2}+ik\right)\,_{2}F_{1}\left(
1,\frac{3}{2}-ik;2;1-\left(\frac{p\,'}{p}\right)^{2}\right)\nonumber\\
&=&\left(\frac{p\,'}{p}\right)^{1-ik}\Gamma\left(\frac{3}{2}-ik\right)\Gamma\left(-\frac{1}{2}+ik\right)
\,_{2}F_{1}\left(
1,\frac{3}{2}-ik;\frac{3}{2}-ik;\left(\frac{p\,'}{p}\right)^{2}\right)\nonumber\\
&&+\left(\frac{p\,'}{p}\right)^{ik}\Gamma\left(\frac{1}{2}-ik\right)\Gamma\left(\frac{1}{2}+ik\right)
\,_{2}F_{1}\left(1,\frac{1}{2}+ik;\frac{1}{2}+ik;\left(\frac{p\,'}{p}\right)^{2}\right)\nonumber\\
&=&\frac{\pi}{\cosh(\pi k)}\left(\left(\frac{p\,'}{p}\right)^{ik}-\left(\frac{p\,'}{p}\right)^{1-ik}\right)\left( 1-\left(\frac{p\,'}{p}\right)^{2}\right)^{-1}.
\end{eqnarray}
We specify that $f_{k}\left(\frac{p}{p\,'}\right)$ is obtained when the substitution $p\rightleftarrows p\,'$ is made.
The second equality from (\ref{ff}) is obtained when Eq.(\ref{hy}) from the Appendix is used.
The above result is valid only when $p\neq p\,'$. We observe that the algebraic argument $\left(\frac{p\,'}{p}\right)^{2}$ in the hypergeometric functions must be considered in the interval $0\leq\left(\frac{p\,'}{p}\right)^{2}<1$. This is the domain of convergence of the hypergeometric functions, because when the algebraic argument becomes equal with $1$, the hypergeometric functions become divergent.
The result of the temporal integrals for $p=p\,'$ can be obtained expressing the integrals with Hankel functions in terms of integrals with Bessel $J$ functions \cite{12,13,20}. The final result is a delta Dirac function $\delta(p-p\,')$ as was shown in \cite{20}. However, when summed the terms with delta Dirac functions cancel out between them for each integral with Hankel functions \cite{20}.

\subsection{The probability}
In this section we will address the physical consequences that emerge from our calculations. The first observation is that
the relevant quantity in our analysis is the transition probability, which is obtained summing the square modulus of the amplitude after the final helicities:
\begin{eqnarray}\label{prf}
\mathcal{P}_{e^-e^+}&=&\frac{1}{2}\sum_{\lambda\lambda\,'}|\mathcal{A}_{e^-e^+} |^{2}=\frac{1}{2}\sum_{\lambda\lambda\,'}\frac{e^{2}}{16\pi^{6}}\frac{1}{|\vec{p}+\vec{p\,'}|^{4}}\nonumber\\
&&\times\biggl\{\frac{\theta(p-p\,')}{p^2}\left[2\left|f_{k}
\left(\frac{p\,'}{p}\right)\right|^{2}-sgn(\lambda\lambda\,')\left(f_{k}^{2}\left(\frac{p\,'}{p}\right)+
f_{k}^{*2}\left(\frac{p\,'}{p}\right)\right)\right]\nonumber\\
&&+\frac{\theta(p\,'-p)}{{p\,'}^{2}}\left[2\left|f_{k}
\left(\frac{p}{p\,'}\right)\right|^{2}-sgn(\lambda\lambda\,')\left(f_{k}^{2}\left(\frac{p}{p\,'}\right)+f_{k}^{*2}\left(\frac{p}{p\,'}\right)\right)\right]\biggl\}\nonumber\\
&&\times|\xi_{\lambda}^{+}(\vec{p}\,)[\vec{\sigma}\cdot(\vec{\mathcal{M}}\times(\vec{p}+\vec{p\,'}))]\eta_{\lambda\,'}(\vec{p\,'})|^{2}.
\end{eqnarray}
The relevant physical parameters that define the probability via the functions $f_{k}\left(\frac{p\,'}{p}\right)\,,f_{k}\left(\frac{p}{p\,'}\right)$ are the ratio $k=m/\omega$ (mass of the fermion/expansion parameter)  and the momenta of the fermions $p\,,p\,'$. This will allow us to consider strong/weak gravitational fields in combination with large/small momenta of the fermions. Note that the influence of the expansion upon the production process in our amplitude/probability enters only via the parameter $k$.

From the final result given in Eqs.(\ref{af}) and (\ref{prf}), we observe that the transition amplitude/probability is nonvanishing only for $p\neq p\,'$. Therefore we can conclude that the momentum conservation law is broken for the process of fermion pair production in the field of a magnetic dipole in de Sitter geometry. It is known that the geometry (\ref{metr}) has spatial translation invariance as an exact symmetry and the momentum is conserved \cite{20}. The presence of the external field breaks this symmetry and this leads to the violation of the momentum conservation on de Sitter geometry. In other words, the momentum is always conserved for the processes like photon emission/absorption by the electron $e^{-}\rightleftarrows e^{-}+\gamma$, pair production/absorption by a single photon $\gamma\rightleftarrows e^{-}+e^{+}$(see, for example, Ref.\cite{21}), where no external fields are present.

Another important aspect is related to the problem of helicity conservation in the pair production process on de Sitter space. It is known that in Minkowski QED, the helicity conservation becomes important in the ultrarelativistic regime $p>>m$ \cite{15,16}. From the probability of transition formula (\ref{prf}), we observe that there are nonvanishing probabilities for production processes that conserve or do not conserve the helicity. In our case the helicity is conserved when $\lambda=-\lambda\,'$, and the helicity conservation law is broken when $\lambda=\lambda\,'$. It is also worthwhile to mention that the helicity conservation law could be broken only for processes that produce massive fermions. All these aspects will be better understood in the next section where we will perform a graphical study of the analytical results.

The total probability for pair production is obtained by integrating (\ref{prf}) after the final momenta $p\,, p\,'$,
\begin{equation}\label{pt}
\mathcal{P}^{tot}_{e^-e^+}=\int \mathcal{P}_{e^-e^+}\,d^3p\,d^3p\,'.
\end{equation}
The integration after the momenta $p\,,p\,'$ in the probability is difficult, since there are terms that contain the momenta ratio at imaginary powers, which are very oscillatory functions. An important observation is that in the probability equation (\ref{prf}), we do not have the delta Dirac function that will allow us to easily perform the momenta integration as in flat space theory. In a future paper we want to approach these integrals that are less studied in literature.

\subsection{Helicity bispinors summation}
The topic of this subsection is the computation of helicity bispinors sums from the probability given in Eq.(\ref{prf}). The two cases of helicity conservation/nonconservation are discussed.
  
The helicity bispinors expressions for $\lambda=\lambda\,'$ and $\lambda=-\lambda\,'$, will be used in our analysis. Then in an orthogonal local frame $\{\vec{e}_i\}$, we take the magnetic moment on the $\vec{e}_{3}$ direction such that $\vec{\mathcal{M}}=\mathcal{M}\,\vec{e}_{3}$ and $\vec{p}\,'=p\,'_{1}\vec{e}_{1}+p\,'_{2}\vec{e}_{2}+p\,'_{3}\vec{e}_{3}\,\,,\vec{p}=p_{1}\vec{e}_{1}+p_{2}\vec{e}_{2}+p_{3}\vec{e}_{3}$. This choice allows us to express the vectorial product in the following form:
\begin{equation}\label{pv}
\vec{\sigma}\cdot(\vec{\mathcal{M}}\times(\vec{p}+\vec{p}\,'))=\mathcal{M}((p_{1}+p_{1}\,')\sigma_{2}-(p_{2}+p_{2}\,')\sigma_{1}).
\end{equation}
From Eq.(\ref{pv}) one can see that the momenta components could be taken in the plane $(1,2)$, since the $p\,_{3},\,p\,_{3}'$ components give no contribution, because of the vectorial product.
Passing to spherical coordinates the electron and positron momenta have the following components $\vec{p}=(p,\alpha,\beta)$ and
${\vec{p}\,}\,'=(p\,',\gamma,\varphi)$, where $\alpha,\, \gamma\in(0,\pi);\,\beta,\, \varphi\in(0,2\pi)$. The usual definition for these components is given below as follows:
\begin{eqnarray}
&&p_{1}=p\sin\alpha\cos\beta;\,\,\,\,\,\, p_{2}=p\sin\alpha\sin\beta;\,\,\,\,\,\,\,p_{3}=p\cos\alpha\nonumber\\
&&p_{1}\,'=p\,'\sin\gamma\cos\varphi;\,\, p_{2}\,'=p\,'\sin\gamma\sin\varphi;\,\,p_{3}\,'=p\,'\cos\gamma.
\end{eqnarray}
Fixing $\alpha=\gamma=\pi/2$, we obtain the situation when the momenta components are in the $(1,2)$ plane. This means that practically we restrict to the polar components of the momenta,
\begin{eqnarray}
&&p_{1}=p\cos\beta;\,\,\,\,\,\, p_{2}=p\sin\beta\nonumber\\
&&p_{1}\,'=p\,'\cos\varphi;\,\, p_{2}\,'=p\,'\sin\varphi.
\end{eqnarray}
The analysis of this case is justified by the fact that the probability of fermion production is maximum when the momenta vectors are perpendicular to the direction of the magnetic dipole as can be seen from (\ref{pv}). If the momenta vectors are in the $(1,2)$ plane, then $\alpha$ is the angle between $\vec{p}$ and $\vec{\mathcal{M}}$ ($\vec{\mathcal{M}}\perp\vec{p}$\,), while $\gamma$ is the angle between $\vec{p}\,\,'$ and $\vec{\mathcal{M}}$ ($\vec{\mathcal{M}}\perp\vec{p}\,'$\,). Another consequence of Eq.(\ref{pv}) is that the fermions could not be generated on the direction of the magnetic field since in this case the probability is zero, as a result of the vectorial product.
Without losing generality we take $\alpha=\gamma=\pi/2$, and in this case we obtain that the angle between momenta of the particle $\vec{p}$ and momenta of the antiparticle $\vec{p}\,\,'$ is just $\beta-\varphi$. The form of the helicity bispinors can be expressed only in terms of polar angles $\beta\,,\varphi$ using the above choice, and the result is given as follows:
\begin{eqnarray}\label{spin}
&&\xi_{1/2}(\vec{p}\,)=\frac{1}{\sqrt{2}}\left(
  \begin{array}{c}
    1 \\
    e^{i\beta} \\
  \end{array}
\right),\,\,\,\,\,\,
\xi_{-1/2}(\vec{p}\,)=\frac{1}{\sqrt{2}}\left(
  \begin{array}{c}
    - e^{-i\beta} \\
   1 \\
  \end{array}
\right)\nonumber\\
&&\eta_{1/2}(\vec{p}\,\,')=\frac{1}{\sqrt{2}}\left(
  \begin{array}{c}
    e^{-i\varphi} \\
    -1 \\
  \end{array}
\right),\,\,\,
\eta_{-1/2}(\vec{p}\,\,')=\frac{1}{\sqrt{2}}\left(
  \begin{array}{c}
   1 \\
    e^{i\varphi} \\
  \end{array}
\right).
\end{eqnarray}

Helicity is conserved when $\lambda=-\lambda\,'$ and in this case two terms will give contribution to the probability given in (\ref{prf}) that corresponds to $\lambda=\frac{1}{2}\,,\lambda'=-\frac{1}{2}$ and $\lambda=-\frac{1}{2}\,,\lambda'=\frac{1}{2}$. Using equation (\ref{spin}), the final result of the helicity bispinors summation when helicity is conserved gives (the term dependent on momenta from the spatial integral is included):
\begin{eqnarray}
&&\,\frac{1}{|\vec{p}+\vec{p}\,'|^{4}}\sum_{\lambda}|\xi_{\lambda}^{+}(\vec{p}\,)[\vec{\sigma}\cdot(\vec{\mathcal{M}}\times(\vec{p}+\vec{p}\,\,'))]\eta_{-\lambda}(\vec{p}\,\,')|^{2}
\nonumber\\
&&=\frac{\mathcal{M}^2}{(p^2+p\,'^2+2pp\,'\cos(\beta-\varphi))^2}(p-p\,')^2(1-\cos(\beta-\varphi)).
\end{eqnarray}

The above equation proves that the final expression of the probability depends on the angle between the momenta vectors. Now taking  $\beta-\varphi=\pi$, we observe that the probability has a maximum. This corresponds to the situation when the electrons and positrons are emitted
perpendicular to the direction of the magnetic field; the momenta vectors are on the same direction and opposite as orientation. For $\beta-\varphi=0$, the probability is vanishing, this being the case when the momenta vectors have the same direction and orientation.

To resume, the electron-positron pair could separate for $\beta-\varphi=\pi$. The helicity conservation law in pair production processes can be one of the physical criteria that could explain the separation between matter and antimatter.
Emission at other angles $\beta-\varphi$ are possible, but these transitions are less probable.

In the opposite case when helicity is not conserved, $\lambda=\lambda\,'=\pm\frac{1}{2}$, the helicity bispinors summation gives:
\begin{eqnarray}
&&\,\frac{1}{|\vec{p}+\vec{p}\,'|^{4}}\sum_{\lambda}|\xi_{\lambda}^{+}(\vec{p}\,)[\vec{\sigma}\cdot(\vec{\mathcal{M}}\times(\vec{p}+\vec{p}\,\,'))]\eta_{\lambda}(\vec{p}\,\,')|^{2}
\nonumber\\
&&=\frac{\mathcal{M}^2}{(p^2+p\,'^2+2pp\,'\cos(\beta-\varphi))^2}(p+p\,')^2(1+\cos(\beta-\varphi)).
\end{eqnarray}
From the above equation one may observe that the probability vanishes for $\beta-\varphi=\pi$, which is the case when the fermions momenta are parallel and have opposite orientations. The interesting situation appears for $\beta-\varphi=0$, when the probability becomes maxim. This is the situation when the momenta vectors have the same direction and orientation, which means that it is very probable for the fermion pair to annihilate into the vacuum.

To conclude, the probability is dependent on the angle between the momenta vectors $\beta-\varphi$, in both helicity conserving/nonconserving cases. Another important result is that, in a magnetic field, the pair of fermions will most probably be generated perpendicular to the field direction in both helicity conserving/nonconserving cases. As a last observation we mention that in an electric field the pairs are emitted parallel with the field direction \cite{23,31,32}.

\section{Graphical results}
\begin{figure}[h!t]
\includegraphics[scale=0.4]{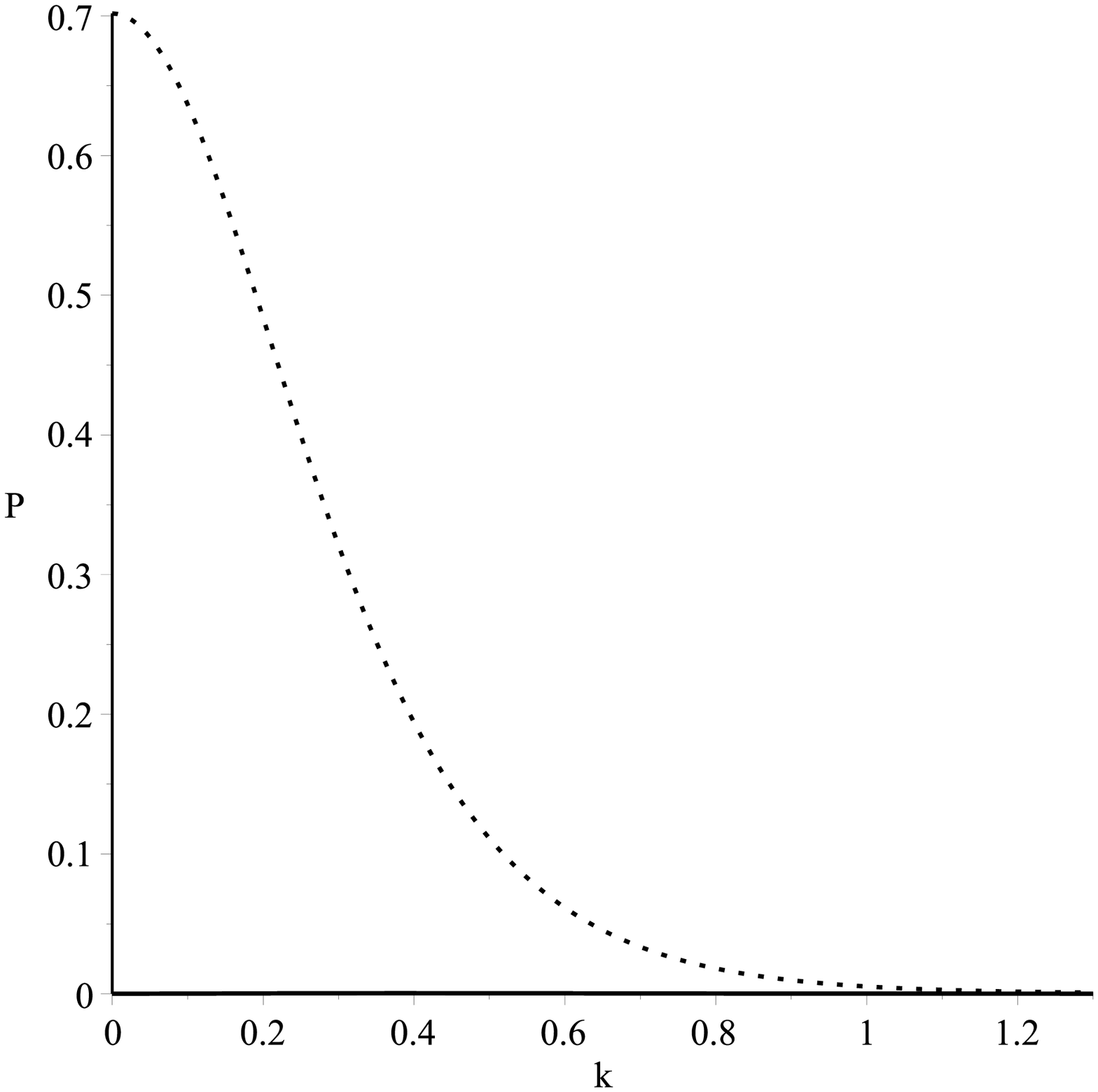}
\includegraphics[scale=0.4]{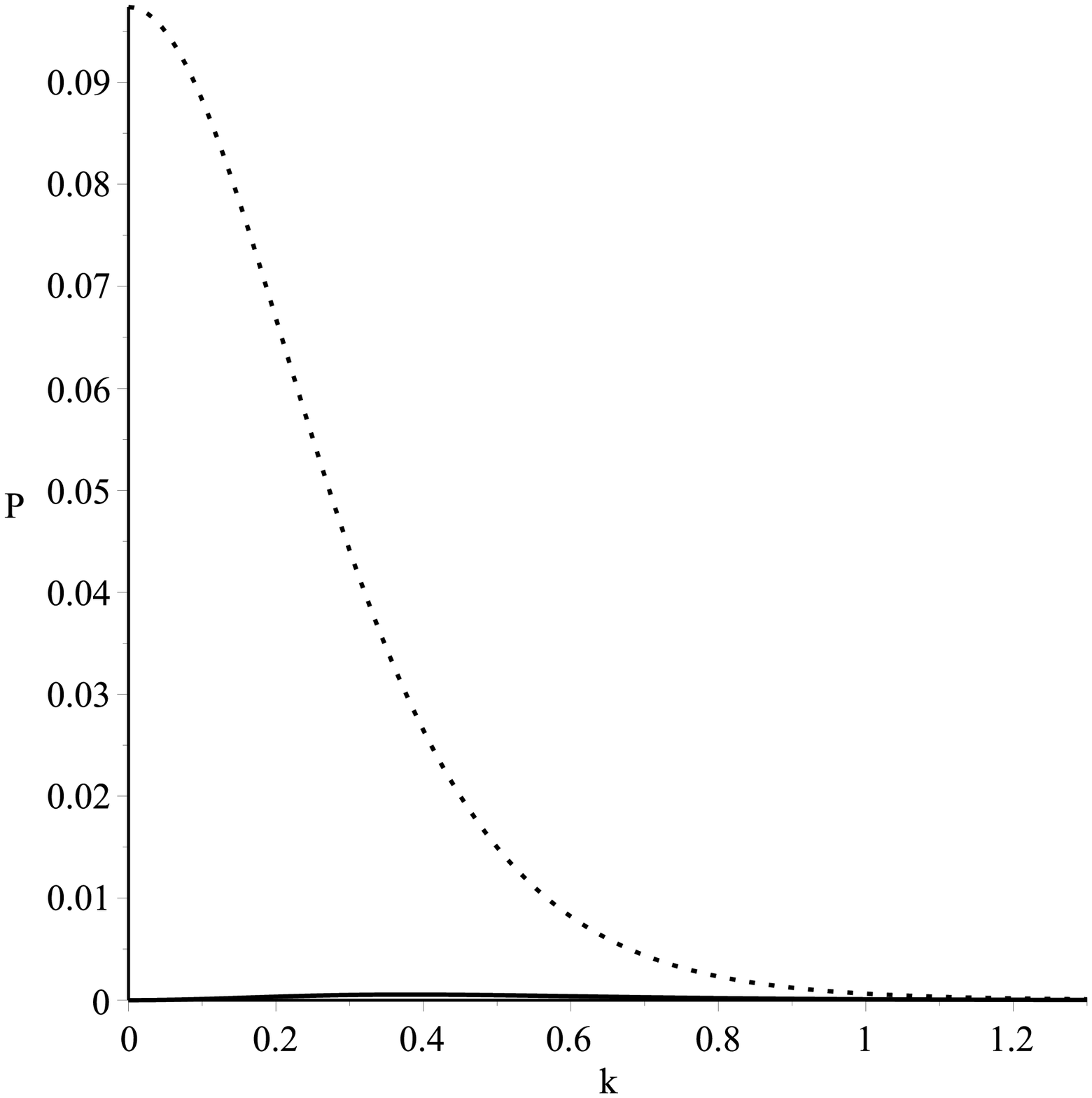}
\caption{Probability as a function of parameter $k$, for $p/p\,'=0.9$ in the left figure and $p/p\,'=0.7$ in right figure. In both cases the point line represents the case of helicity conservation.}
\label{f1}
\end{figure}

\begin{figure}[h!t]
\includegraphics[scale=0.4]{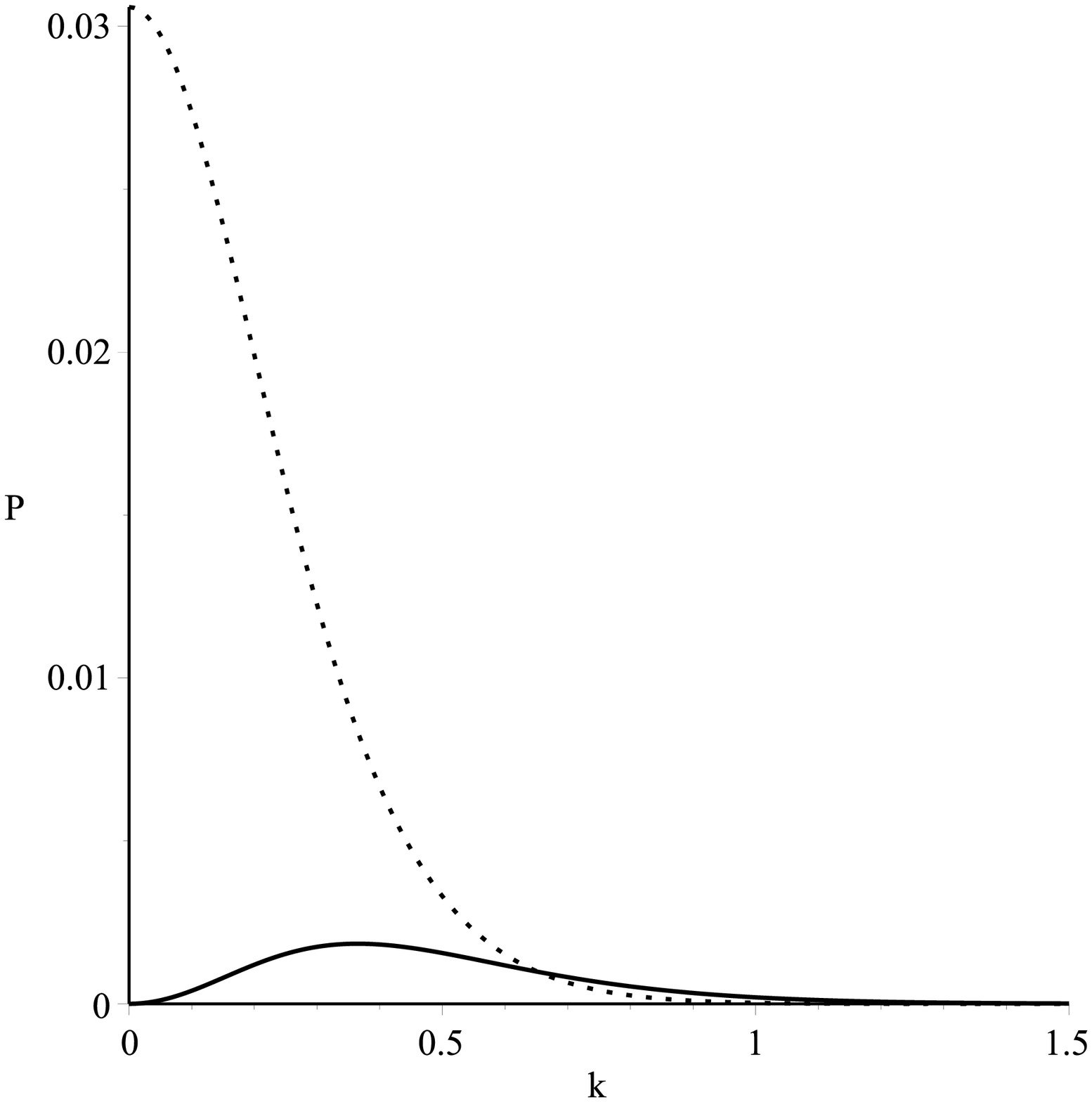}
\includegraphics[scale=0.4]{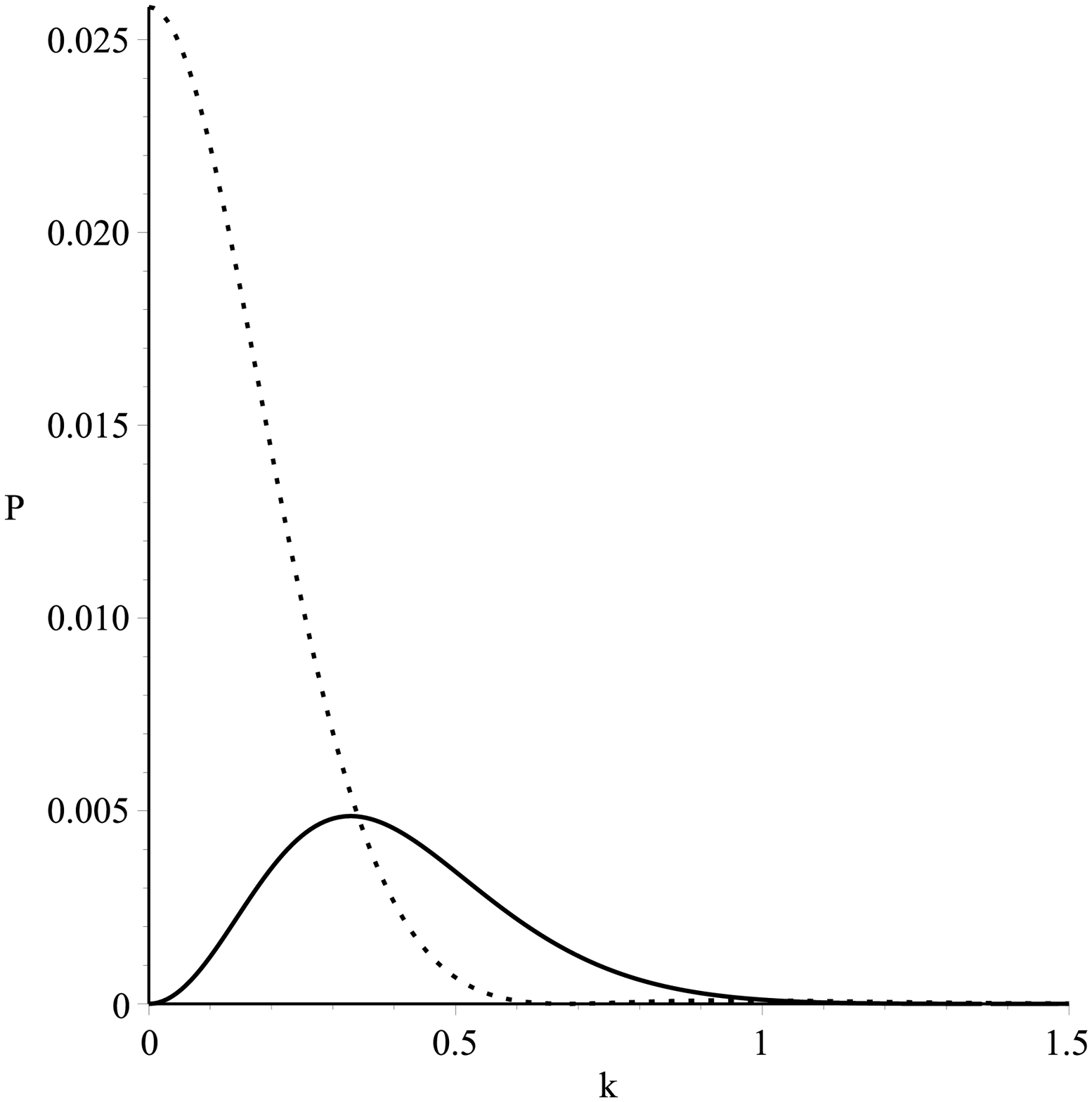}
\caption{Probability as a function of parameter $k$, for $p/p\,'=0.3$ in the left figure and $p/p\,'=0.1$ in the right figure. The point line represents the case of helicity conservation and the solid line the case when helicity is not conserved.}
\label{f2}
\end{figure}

\begin{figure}[h!t]
\includegraphics[scale=0.4]{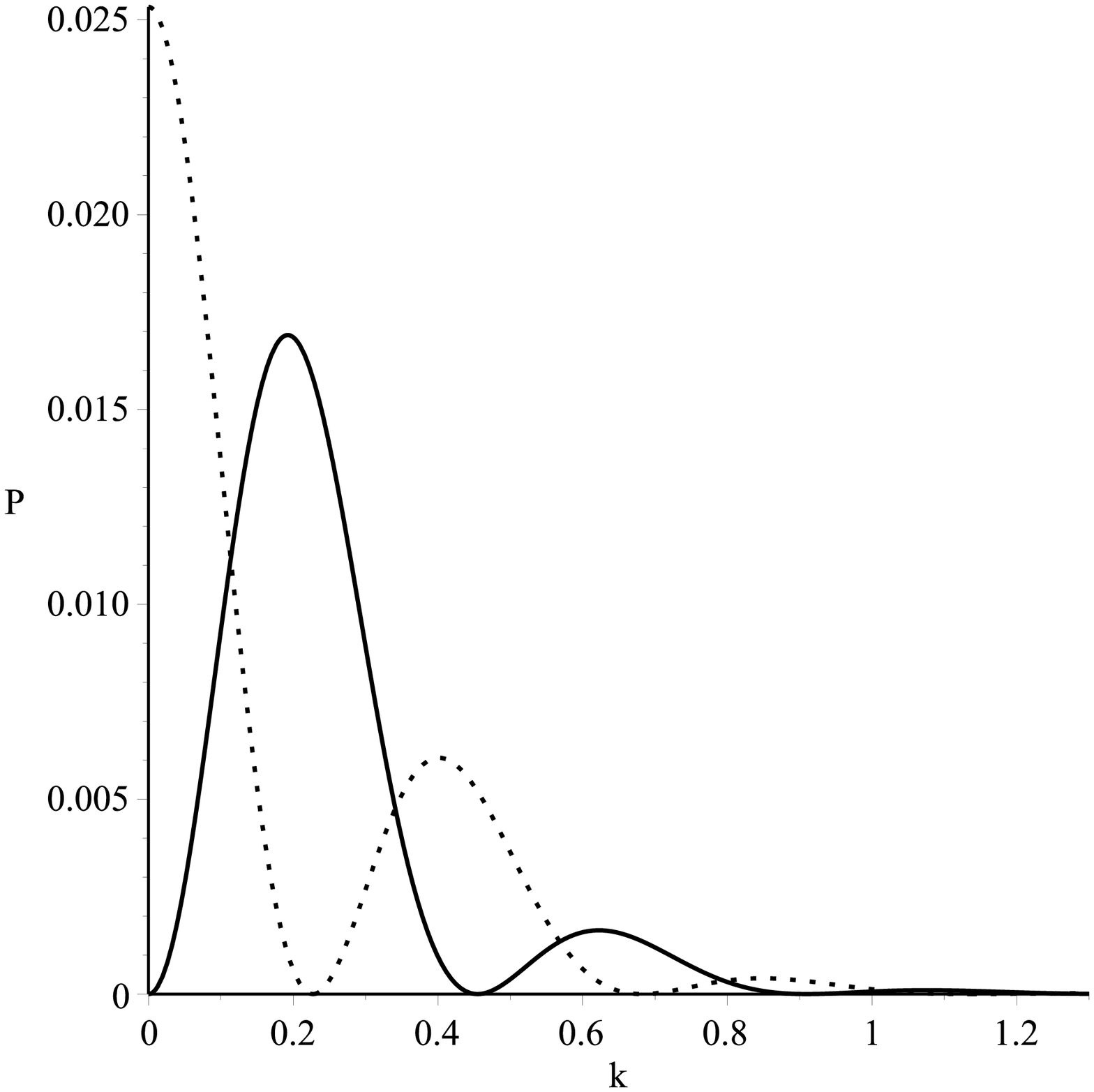}
\includegraphics[scale=0.4]{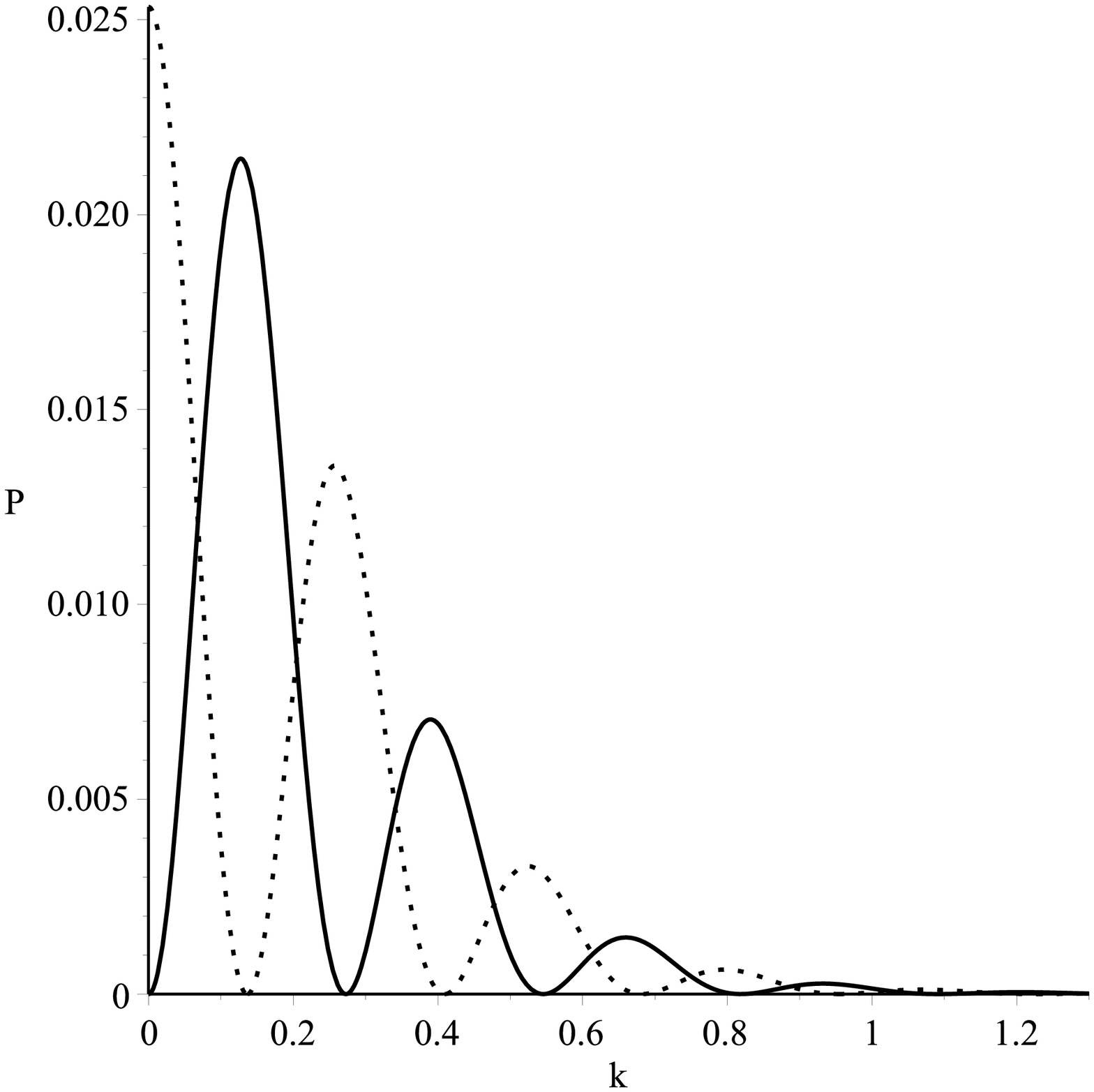}
\caption{Probability as function of parameter $k$, for $p/p\,'=0.001$ in the left figure and $p/p\,'=0.00001$ in the right figure. The point line represents the case of helicity conservation and the solid line the case when helicity is not conserved.}
\label{f3}
\end{figure}

In this section we will perform a graphical analysis of the probability because its dependence on the physical parameters $k,\,p\,,p\,'$ is complicated and only in this way could one better understand the significance of the analytical results. This will allow us to consider strong/weak gravitational fields in combination with small/large values of the momenta modulus $p\,,p\,'$. Also in this section we will study the probability dependence in terms of the angle between the momenta of the fermions $\beta-\varphi$, when the parameters $k,\,p,\,p\,'$ are given. Using the results of the previous section the final expression for the probability reads (both cases of helicity conservation/nonconservation are considered):
\begin{eqnarray}\label{prf1}
\mathcal{P}_{e^-e^+}&=&\frac{e^{2}\mathcal{M}^2}{16\pi^{6}(p^2+p\,'^2+2pp\,'\cos(\beta-\varphi))^2}\nonumber\\
&&\times\biggl\{\frac{\theta(p-p\,')}{p^2}\left[2\left|f_{k}
\left(\frac{p\,'}{p}\right)\right|^{2}\pm\left(f_{k}^{2}\left(\frac{p\,'}{p}\right)+
f_{k}^{*2}\left(\frac{p\,'}{p}\right)\right)\right]\nonumber\\
&&+\frac{\theta(p\,'-p)}{{p\,'}^{2}}\left[2\left|f_{k}
\left(\frac{p}{p\,'}\right)\right|^{2}\pm\left(f_{k}^{2}\left(\frac{p}{p\,'}\right)+f_{k}^{*2}\left(\frac{p}{p\,'}\right)\right)\right]\biggl\}\nonumber\\
&&\times\left\{
\begin{array}{cll}
(p-p\,')^2(1-\cos(\beta-\varphi))&{\rm for}&\lambda=-\lambda'\\
(p+p\,')^2(1+\cos(\beta-\varphi))&{\rm for}&\lambda=\lambda'
\end{array}\right.\label{ij}.
\end{eqnarray}
From this equation we observe that the probability has a simple quadratic dependence in terms of the magnetic moment $\mathcal{M}$. Further, we plot the above probability as a function of parameter $k$ for given values of the momenta $p,\,p\,'$ and fixed angles (the software used for obtaining our graphs is MAPLE). In the helicity conserving case we fix $\beta-\varphi=\pi\,$ since in this case the probability becomes maxim, while in the helicity nonconserving case, the probability becomes maxim if we fix $\beta-\varphi=0$. These results are presented in Figs.(\ref{f1})-(\ref{f3}). In our graphs the numerical values for the dipole moment are taken in the interval $\mathcal{M}\in(1,3)$.

From Figs.(\ref{f1})-(\ref{f3}), we observe that the probability of fermion production is significant only for small values of the parameter $k$. This result shows that the phenomenon of fermion production is important only in strong gravitational fields or equivalently when the expansion factor $\omega$ is large comparatively with the fermion mass. In other words the graphical results prove that the phenomenon of pair production is significant only in the early Universe. This conclusion is in accordance with the important results obtained by Parker \cite{25,26,27}, which also proves that the phenomenon of particle production was important only in the early Universe. When the parameter $k\sim1$, the probabilities decrease rapidly and approach to zero, in both helicity conserving/nonconserving cases.

\ \

\ \

\ \

For large $k$ the probability for pair production reduces to zero (see Figs.(\ref{f1})-(\ref{f3})), this being the first confirmation that we obtain the Minkowski limit, which corresponds to $k=\infty$.

In the helicity nonconserving case we obtain that the probability of pair production vanishes for massless fermions (see Figs.(\ref{f1})-(\ref{f3})). This result is expected since the helicity conservation law could be broken only for massive fermions. The graphical results show that the probability of fermion production in the helicity nonconserving case is negligible comparatively with the probability of production in the helicity conserving case, when the ratio of the momenta is close to unity (see Fig. (\ref{f1})). As the ratio of the momenta becomes smaller, $p/p\,'<<1$, the probability of pair production in the helicity nonconserving case becomes important (see Figs.(\ref{f2})-(\ref{f3})). From our graphs it is clear that the most probable processes are those that conserve the helicity and have the ratio of the momenta close to unity. To conclude when the momenta modulus are close as value ($p\sim p\,'$), there is a tendency for helicity conservation while the helicity nonconservation is present when the momenta modulus are different as value ($p>> p\,'$). Our graphs shows that the processes with helicity conservation will be dominant in the case of pair production in the field of a magnetic dipole in de Sitter space.

Also from Figs.(\ref{f1})-(\ref{f3}), one may observe that the oscillatory behavior of the probabilities becomes more pronounced as the ratio of the momenta $p/p\,'$ becomes smaller. This behavior originates from the probability equation, which is proportional with terms of the form $\left(\frac{p\,'}{p}\right)^{\pm2ik}$. Another interesting aspect related to the graphs is that fermions are created up to $m/\omega\sim1$, and in this interval the probabilities in both helicity conserving and nonconserving cases have a series of minima and maxima that decrease as $m/\omega$ increase. If we increase the dipole moment modulus $\mathcal{M}$, a graphical analysis shows that the probabilities become bigger and will have nonvanishing values in the same interval $m/\omega\in(0,1)$ as we mentioned above. So our analysis was done for a small magnetic dipole moment, placed in strong gravitational fields of the early Universe. It is clear now that our results are valid only for the large expansion phase, and in the Minkowski limit the probability is zero.

In what follows we will analyze the dependence of the probability in terms of the angle between the momenta vectors $\beta-\varphi$, for given values of the parameter $k$ and different values of the momenta ratio $p/p\,'$. The two situations when the helicity is conserved or not conserved will be included in our analysis.
\begin{figure}[h!t]
\includegraphics[scale=0.4]{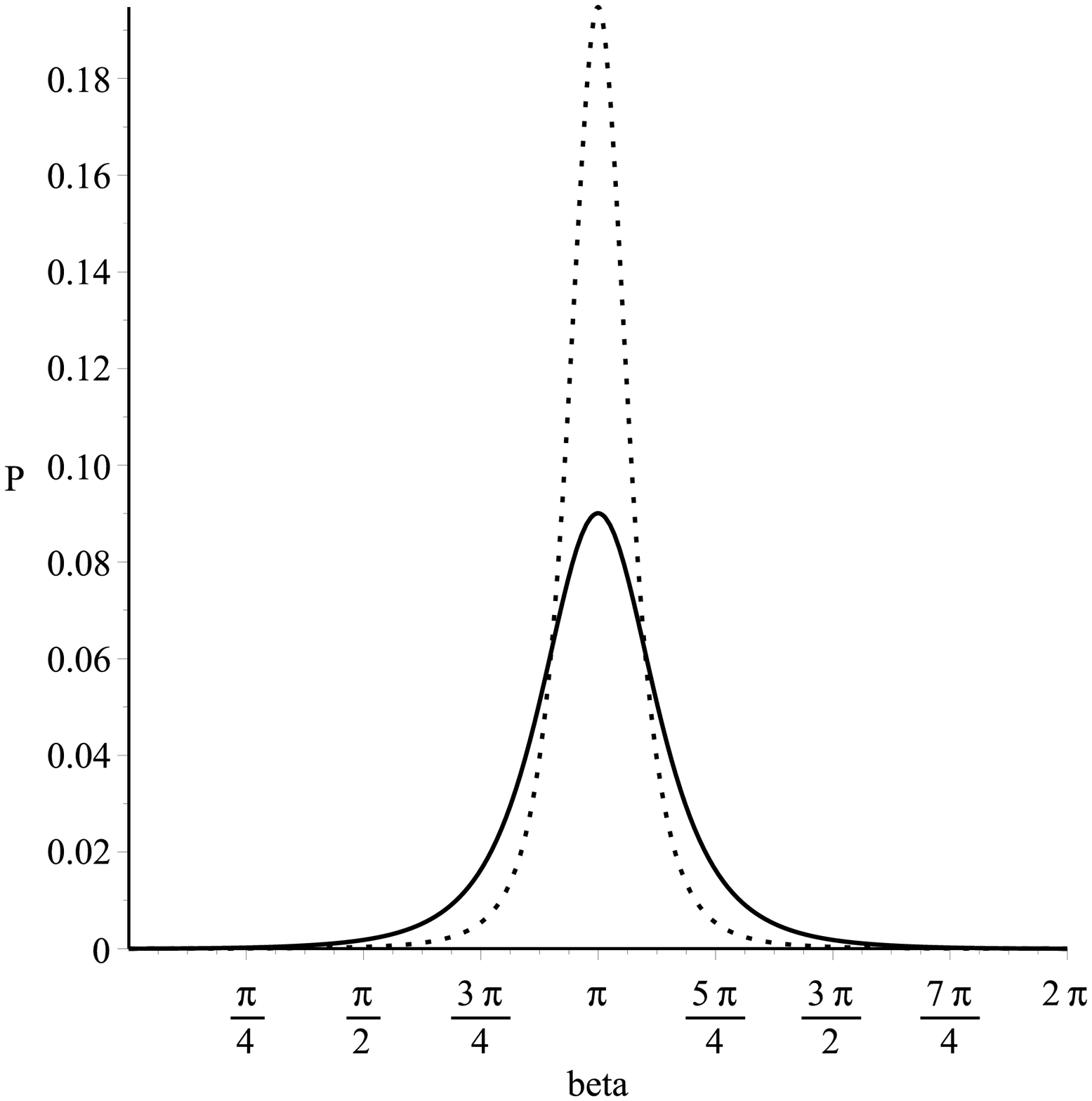}
\includegraphics[scale=0.4]{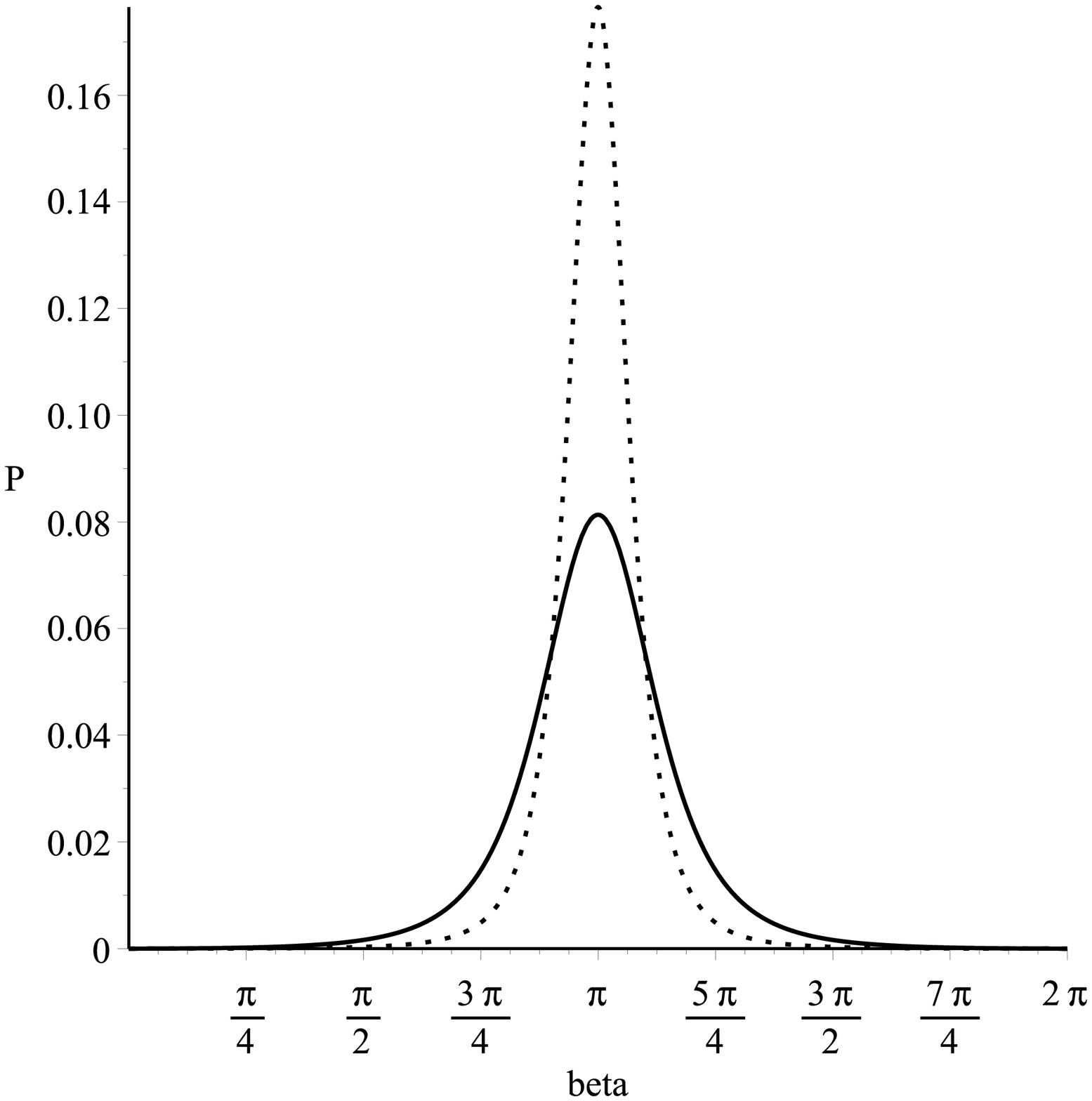}
\caption{Probability as a function of angle $\beta-\varphi$ when helicity is conserved, for $p/p\,'=0.7$ point line $p/p\,'=0.5$ solid line. Left figure is for $k=0.001$ and the right figure for $k=0.1$.}
\label{f4}
\end{figure}

\begin{figure}[h!t]
\includegraphics[scale=0.4]{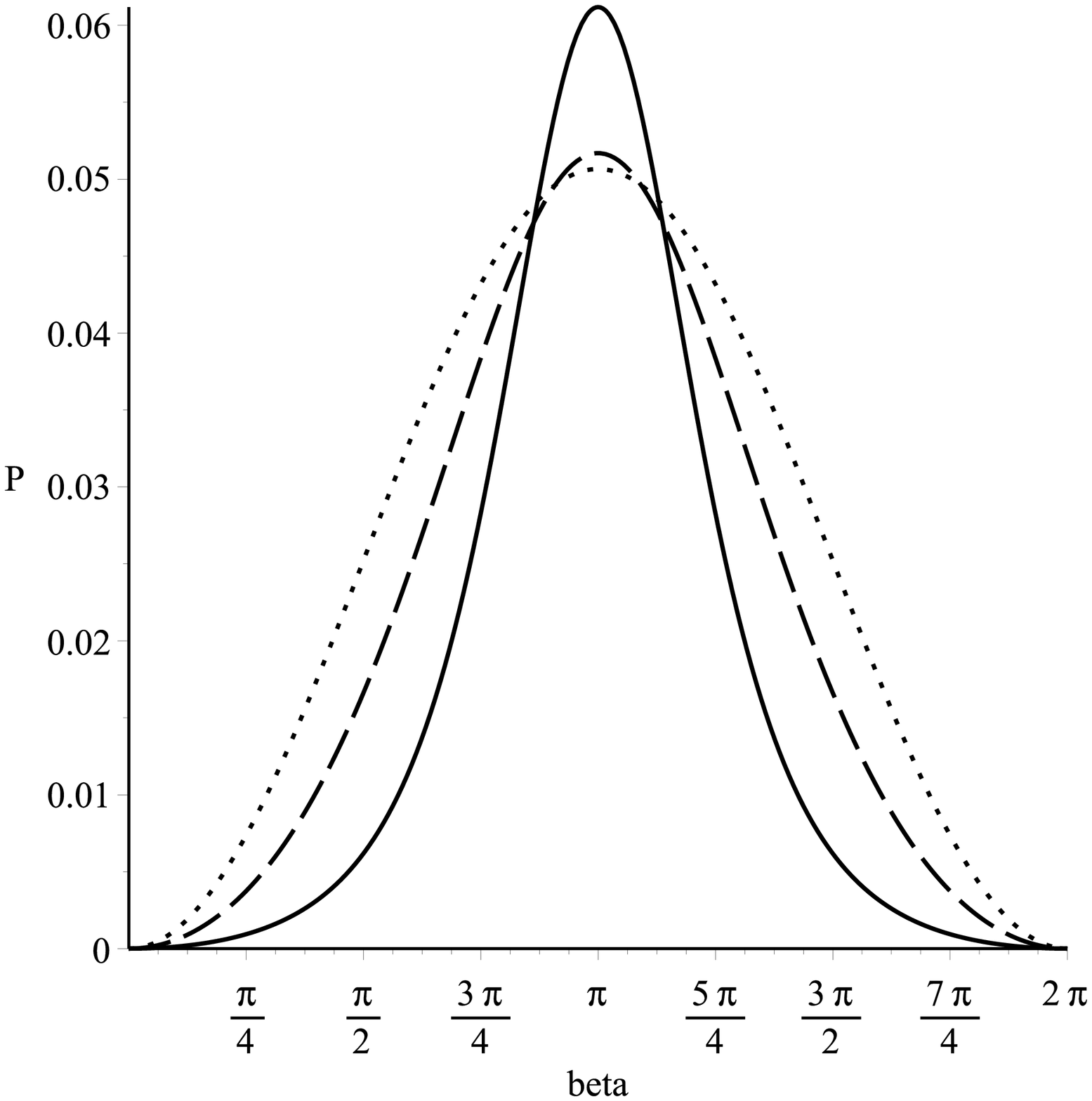}
\includegraphics[scale=0.4]{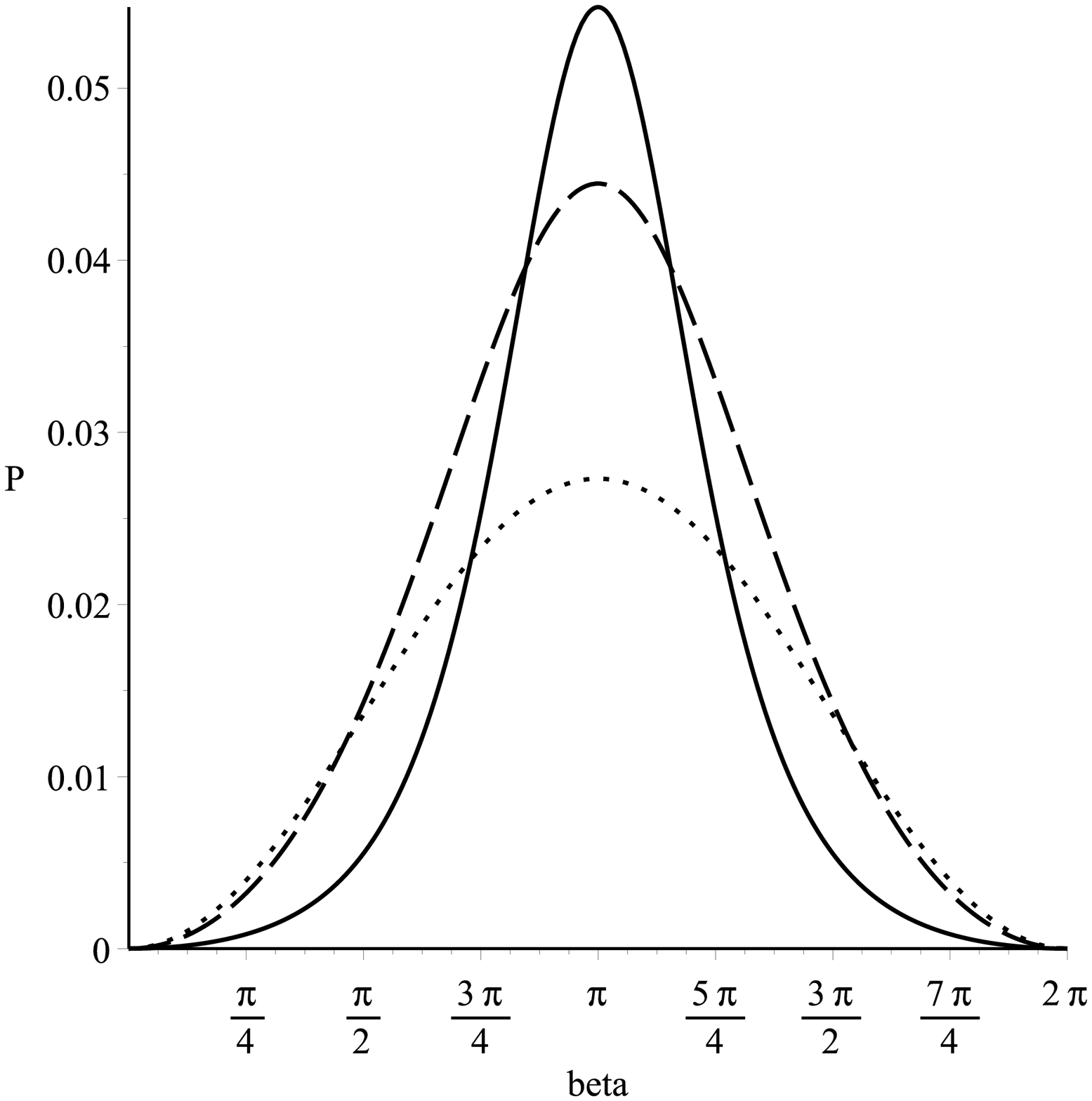}
\caption{Probability as a function of angle $\beta-\varphi$ when helicity is conserved, for $p/p\,'=0.001$ point line $p/p\,'=0.3$ solid line and $p/p\,'=0.1$ for dashed line. Left figure is for $k=0.001$ and the right figure for $k=0.1$.}
\label{f5}
\end{figure}

\begin{figure}[h!t]
\includegraphics[scale=0.4]{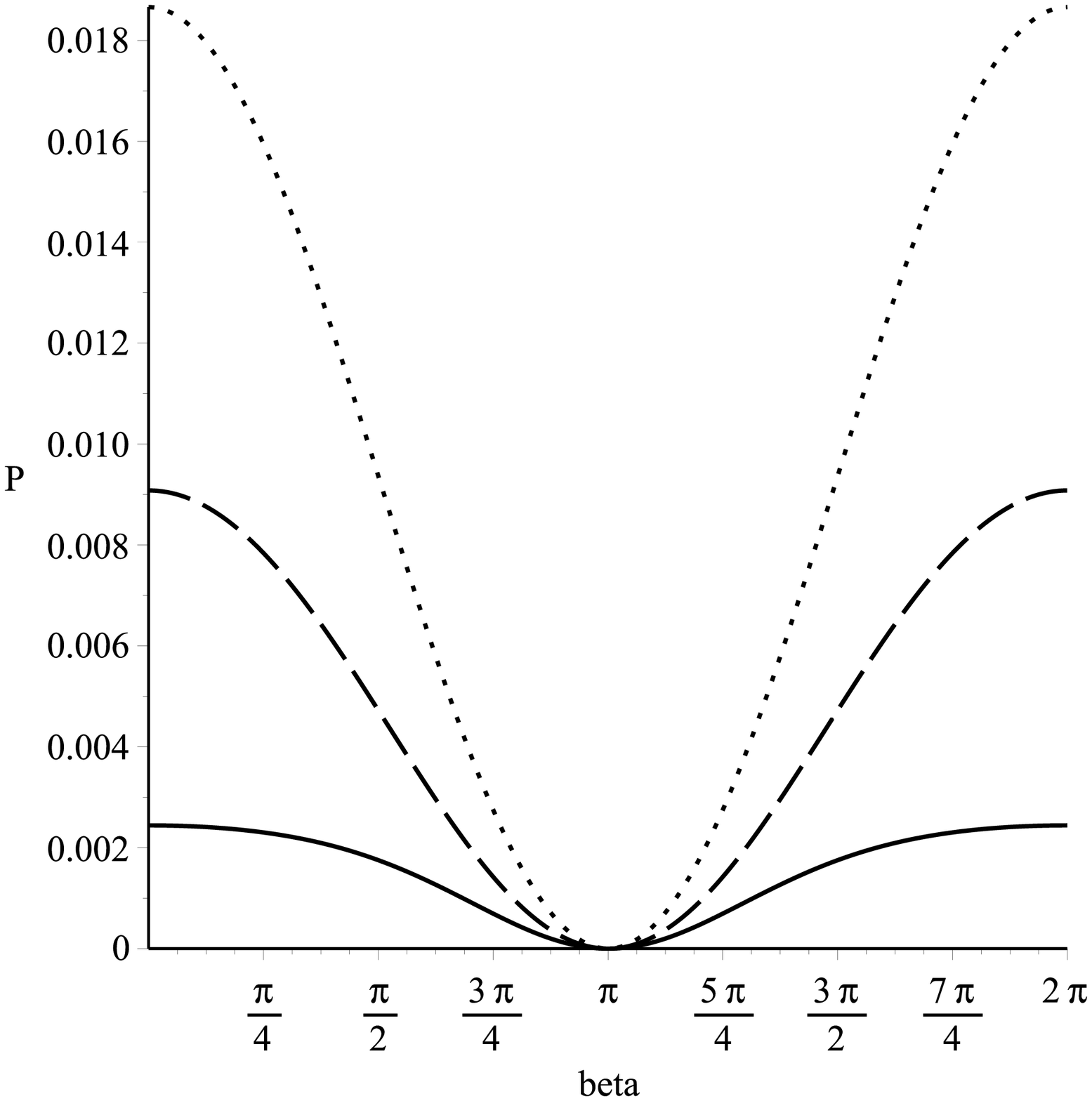}
\includegraphics[scale=0.4]{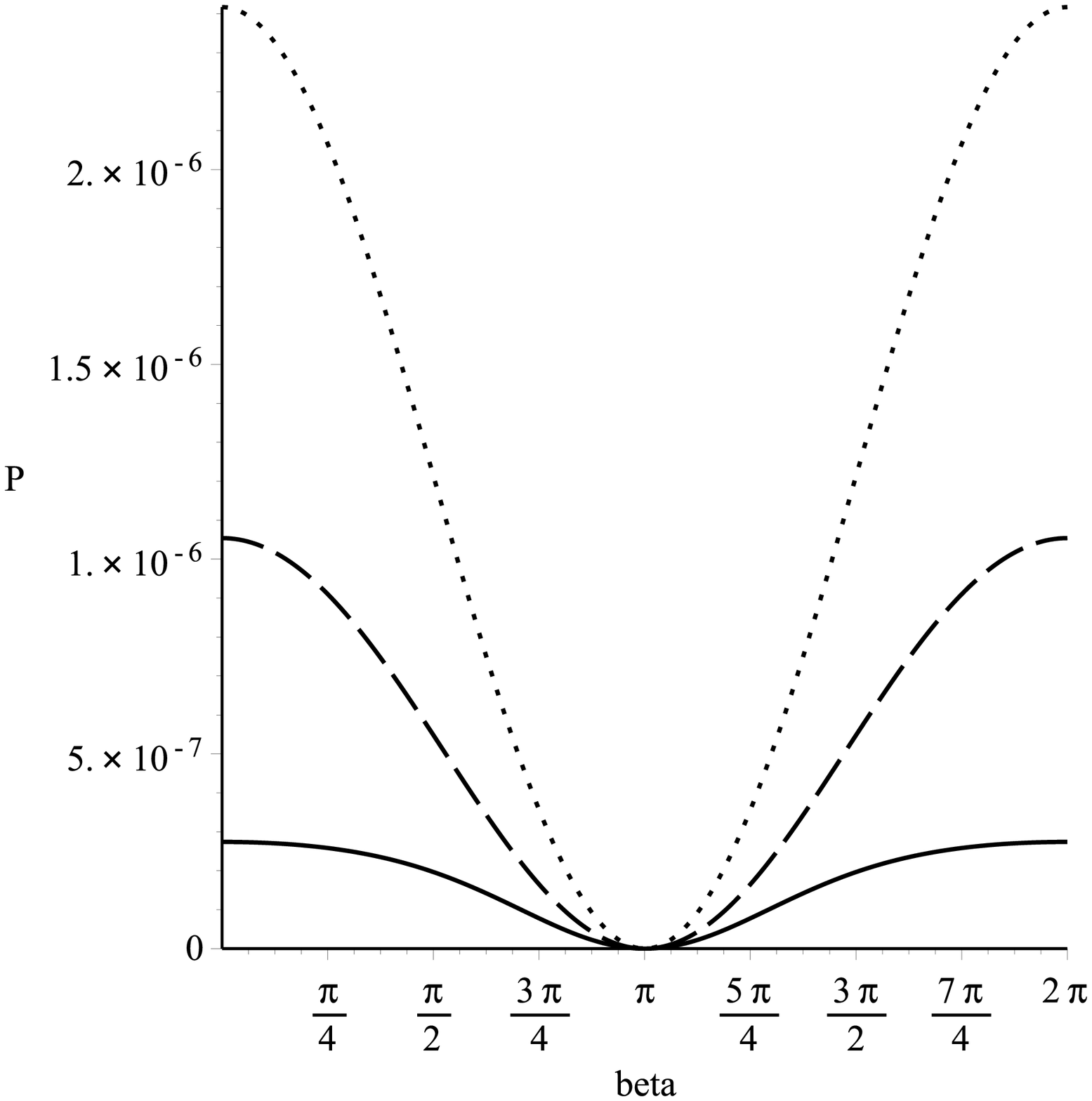}
\caption{Probability as a function of angle $\beta-\varphi$ when helicity is not conserved, for $p/p\,'=0.001$ point line $p/p\,'=0.1$ solid line and $p/p\,'=0.01$ for dashed line. Left figure is for $k=0.1$ and the right figure for $k=0.001$.}
\label{f6}
\end{figure}

\begin{figure}[h!t]
\includegraphics[scale=0.4]{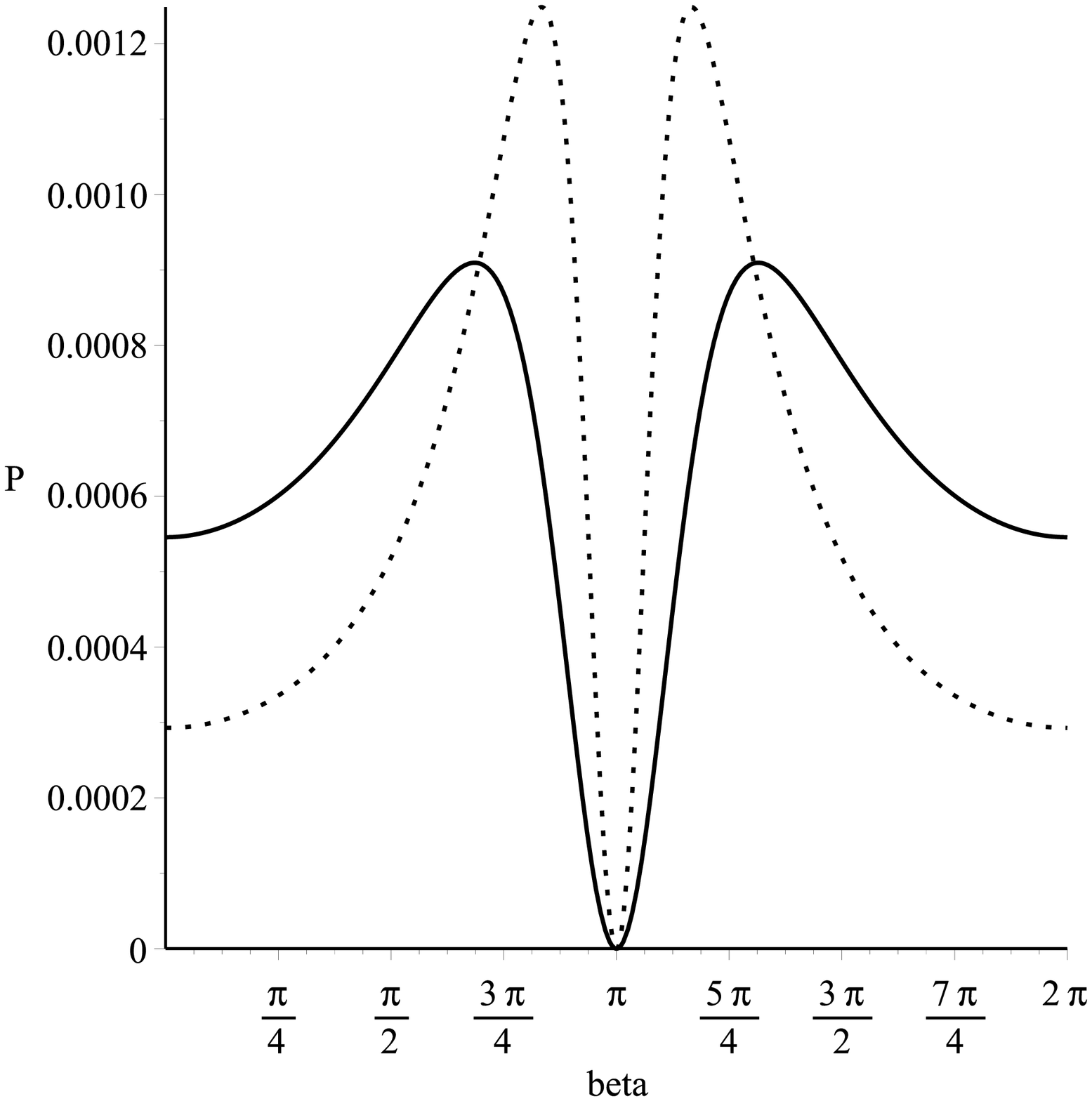}
\includegraphics[scale=0.4]{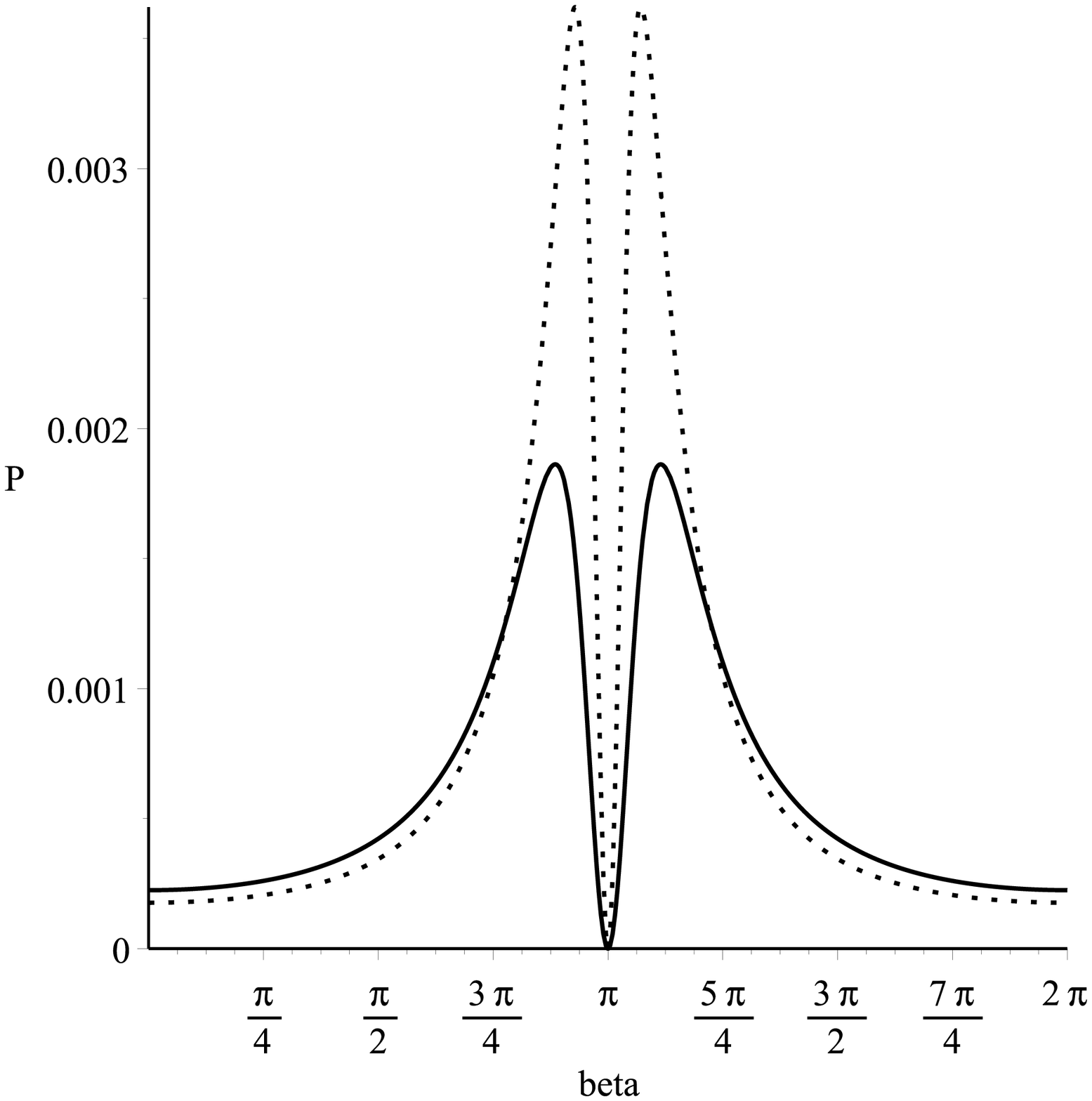}
\caption{Probability as a function of angle $\beta-\varphi$ when helicity is not conserved, in left figure $p/p\,'=0.6$ point line and $p/p\,'=0.4$ solid line, while in the right figure $p/p\,'=0.8$ point line and $p/p\,'=0.7$ solid line. In both figures $k=0.1$.}
\label{f7}
\end{figure}

\ \

\ \

\ \

In Fig.(\ref{f4}), we observe that the probability of pair production in the helicity conserving case is nonvanishing only when the angle between momenta vectors is close to $\pi$. As the ratio $p/p\,'$ approaches unity, we observe that the probability has more pronounced maxima in $\beta-\varphi=\pi$. The graphical result obtained Fig.(\ref{f4}), confirm the observations made in the previous section and show that in the helicity conserving case the matter and antimatter could separate, since the momenta vectors of the pair are opposite at orientation and are aligned in the same direction.

In Fig.(\ref{f5}) we present the probability dependence of the angle in the helicity conserving case, with the specification that the small values of the ratio $p/p\,'$ were considered. Here one can observe that the probability maximum at $\beta-\varphi=\pi$ is not so pronounced and there are nonvanishing probabilities for pair generation even for smaller values of the angle between momenta vectors $\beta-\varphi$. The variation with parameter $k$ shows that as $k\rightarrow0$, the probability increases since the gravity becomes stronger (or equivalently the mass approaches zero).

In Figs.(\ref{f6})-(\ref{f7}) we present the angular dependence of the probability of pair production in the case when helicity is not conserved.
From Fig.(\ref{f6}) we see that when the momenta ratio is small, the probability has a maximum at $\beta-\varphi=0$ and is zero for $\beta-\varphi=\pi$. When the parameter $k$ is close to zero, the probability of pair production in the helicity nonconserving case becomes very small. This is expected since at zero mass the probability vanishes in the helicity nonconserving case, as we can observe from our plots in terms of parameter $k$ (see Figs.(\ref{f2})-(\ref{f3})). Fig.(\ref{f7}) gives the probability dependence of the angle $\beta-\varphi$ when the momenta ratio $p/p\,'$ is close to unity. From Fig.(\ref{f7}) we see that the probability has two maxims for angles close to $\pi$, and it is worth to mention that these maxims increase and become more closer to $\pi$ as the ratio $p/p\,'$ approaches unity. This proves that when  $p/p\,'$ is close to unity, the maximum value of the probability is shifted from small $\beta-\varphi\sim0$ to large angles close to $\pi$, even when the helicity is not conserved. From here we can draw the conclusion that the pair could separate even in the helicity nonconserving case when the momenta moduli are close in value, but the probability for such a process is much smaller comparatively with the helicity conserving case presented in Fig.(\ref{f4}).

\begin{figure}[h!t]
\includegraphics[scale=0.4]{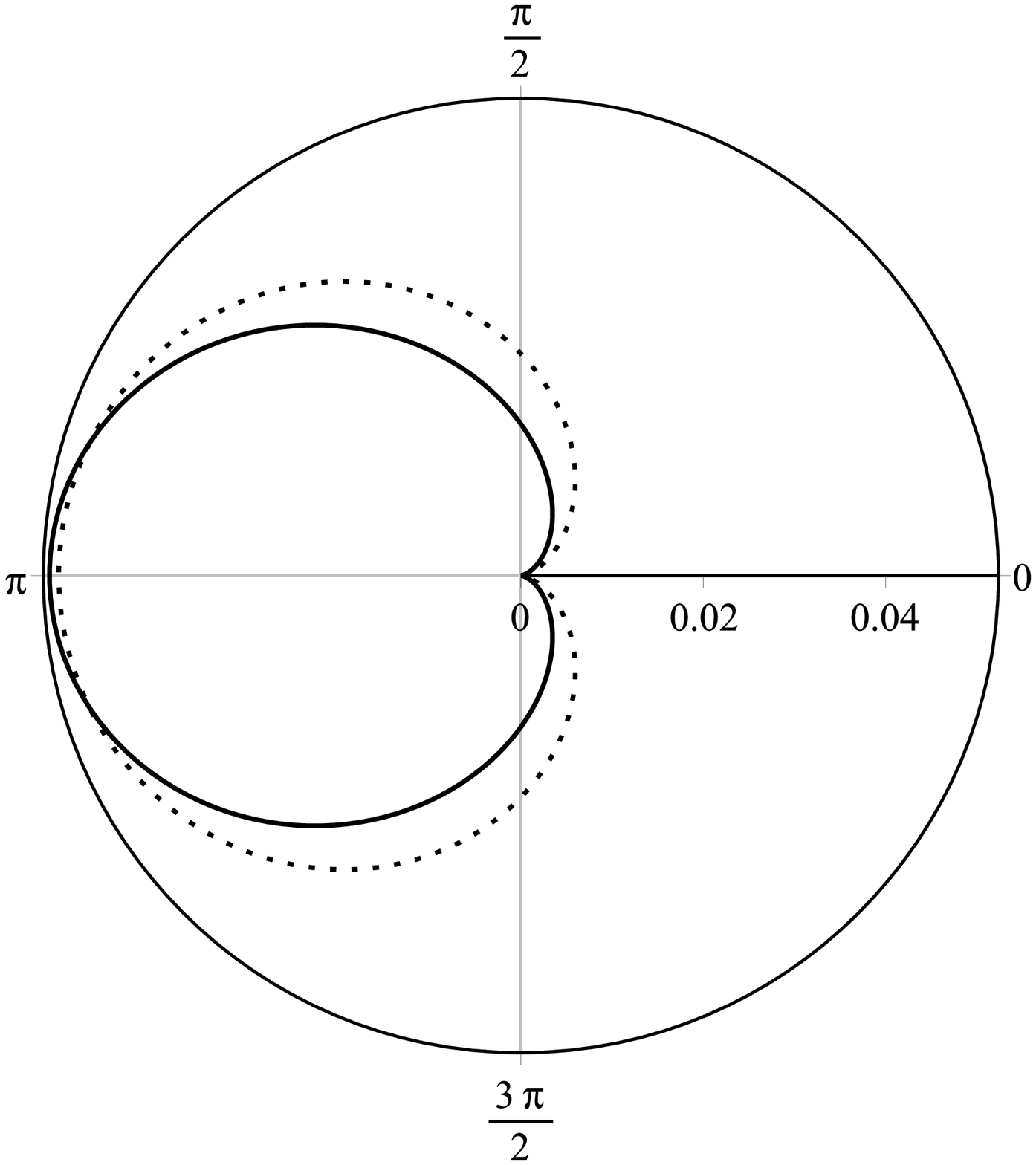}
\includegraphics[scale=0.4]{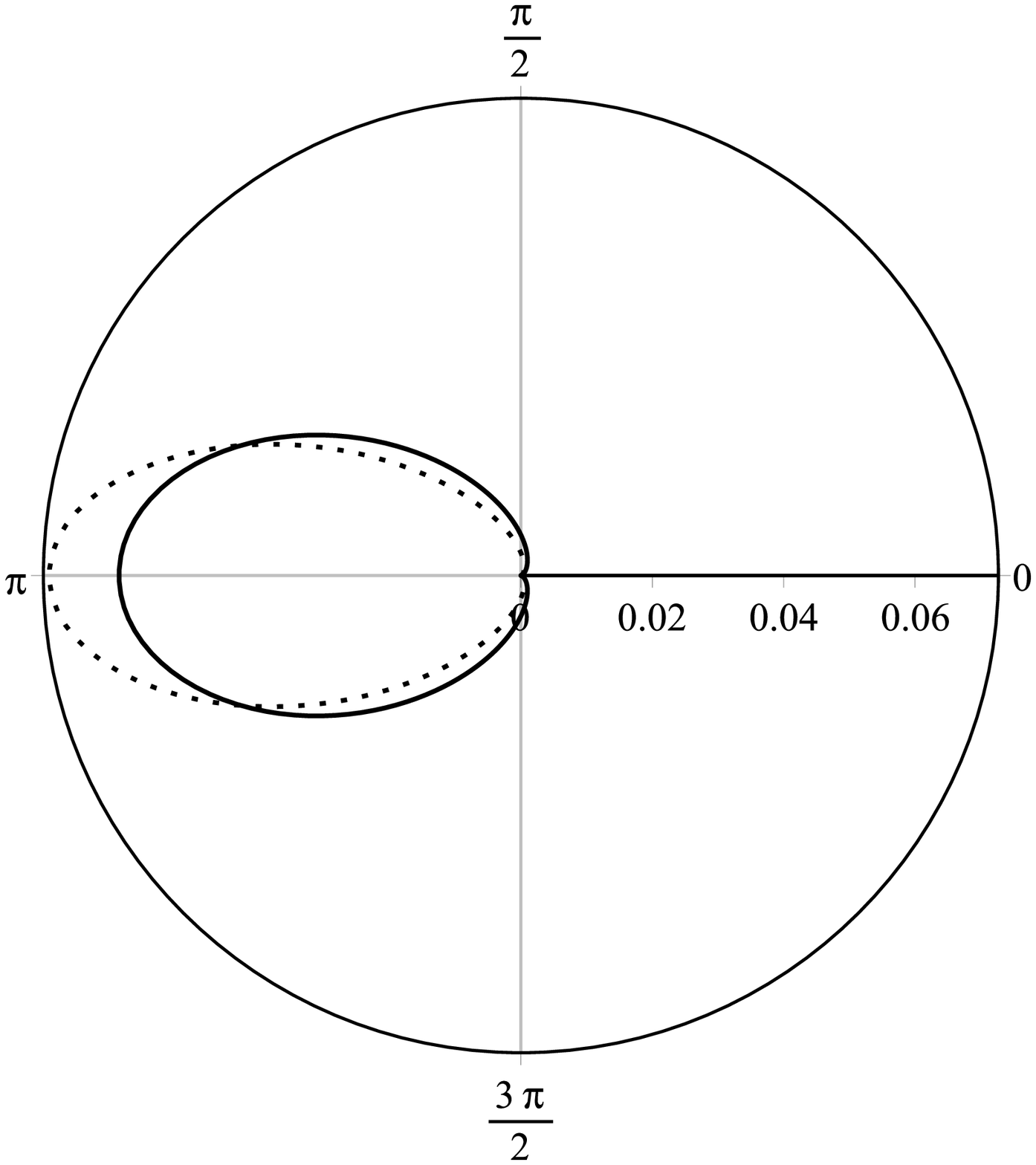}
\caption{Polar plot of the probability as a function of angle $\beta-\varphi$, in the case of helicity conservation. Parameter $k=0.001$ in both figures. In the left figure $p/p\,'=0.1$ solid line and $p/p\,'=0.01$ the point line, while in the right figure $p/p\,'=0.3$ solid line and $p/p\,'=0.4$ the point line.}
\label{f8}
\end{figure}

\begin{figure}[h!t]
\includegraphics[scale=0.4]{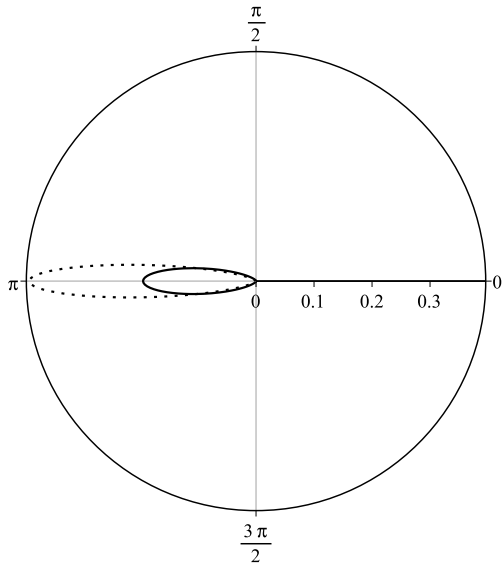}
\caption{Polar plot of the probability as a function of angle $\beta-\varphi$, in the case of helicity conservation for $k=0.001$. The solid line is for $p/p\,'=0.7$ and the point line for $p/p\,'=0.8$.}
\label{f9}
\end{figure}

\begin{figure}[h!t]
\includegraphics[scale=0.4]{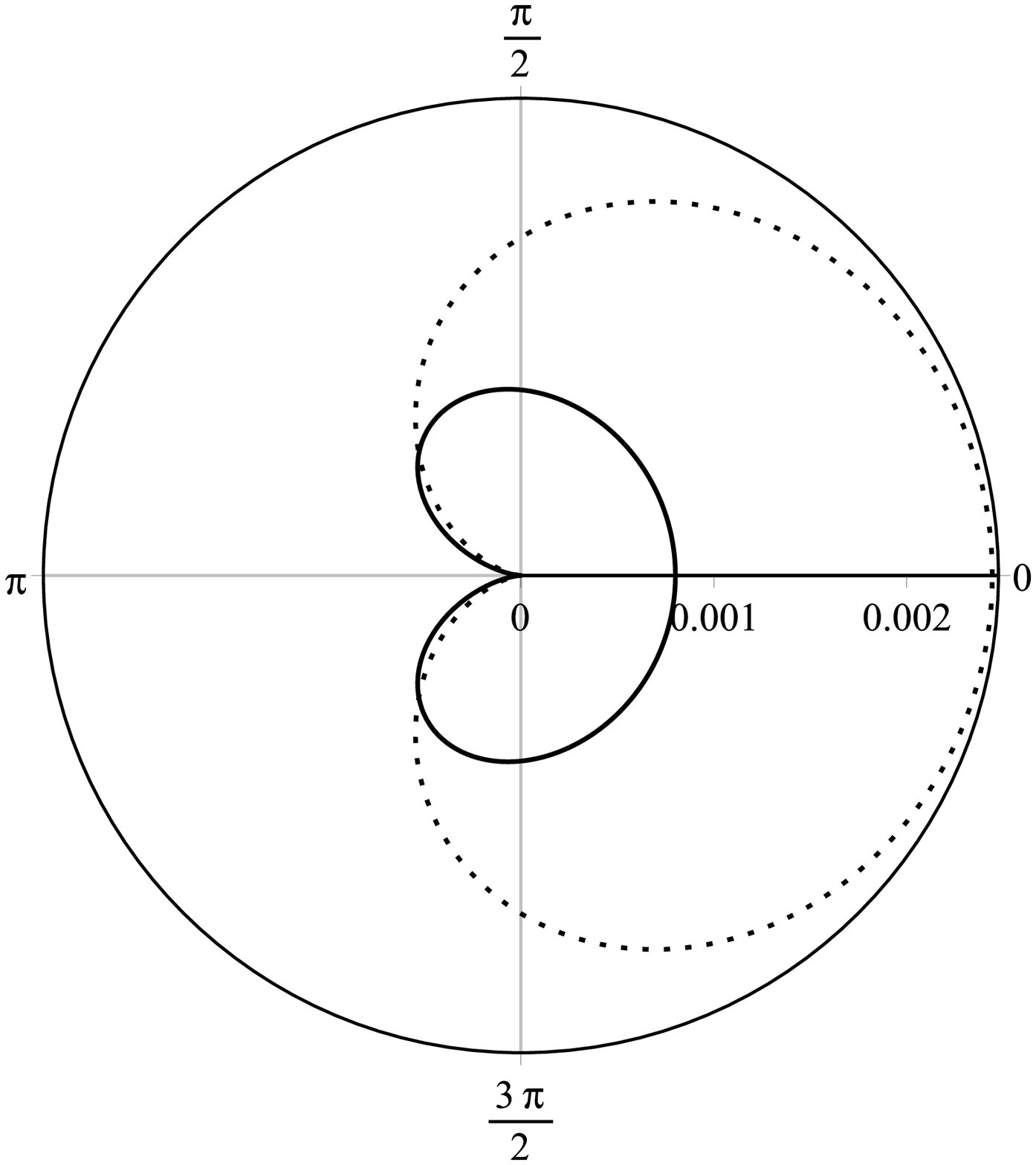}
\includegraphics[scale=0.4]{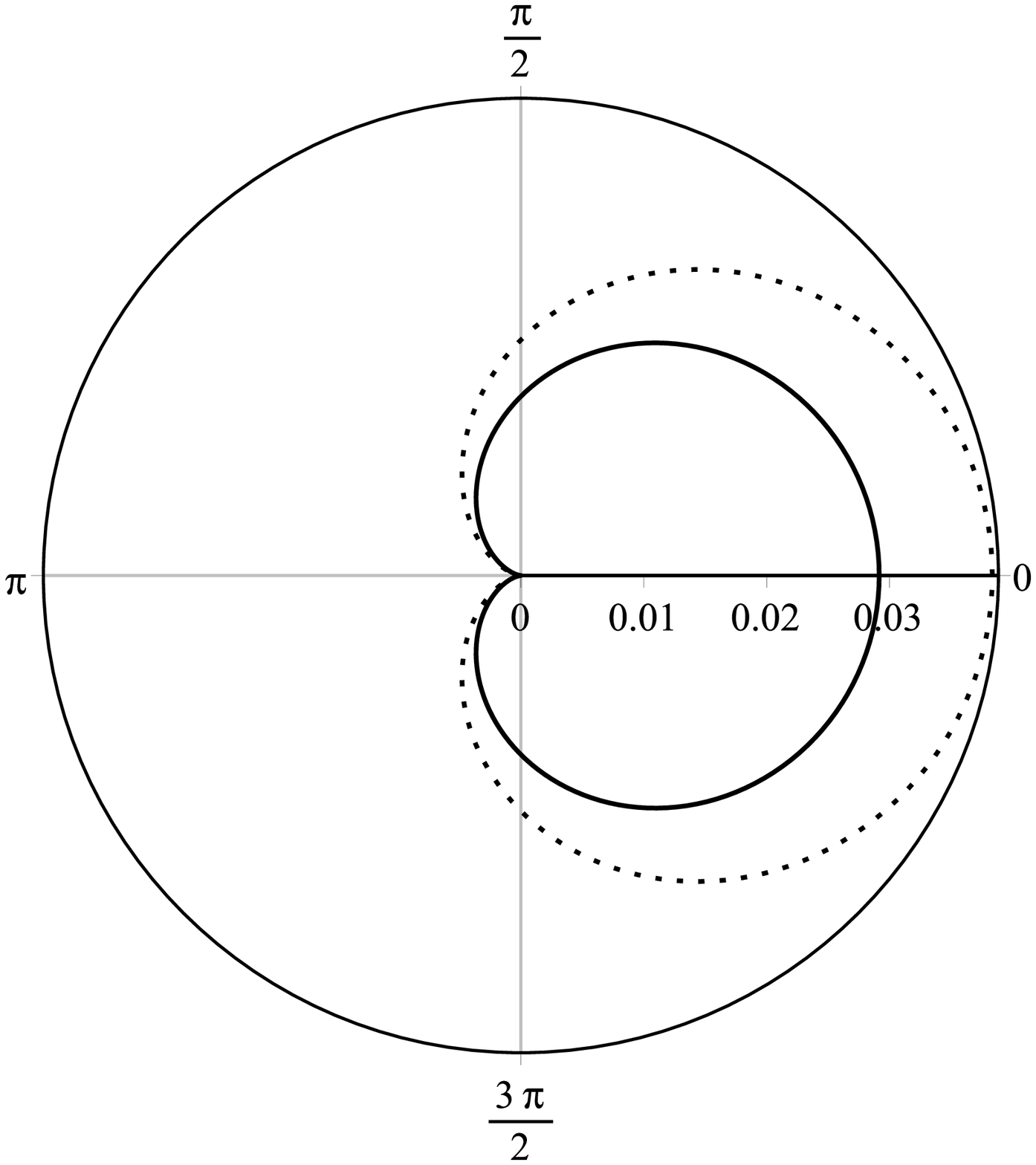}
\caption{Polar plot of the probability as a function of angle $\beta-\varphi$, in the case when helicity is not conserved. Parameter $k=0.1$ in both figures. In the left figure $p/p\,'=0.3$ solid line and $p/p\,'=0.1$ the point line, while in the right figure $p/p\,'=0.0001$ solid line and $p/p\,'=0.00001$ the point line.}
\label{f10}
\end{figure}

\begin{figure}[h!t]
\includegraphics[scale=0.4]{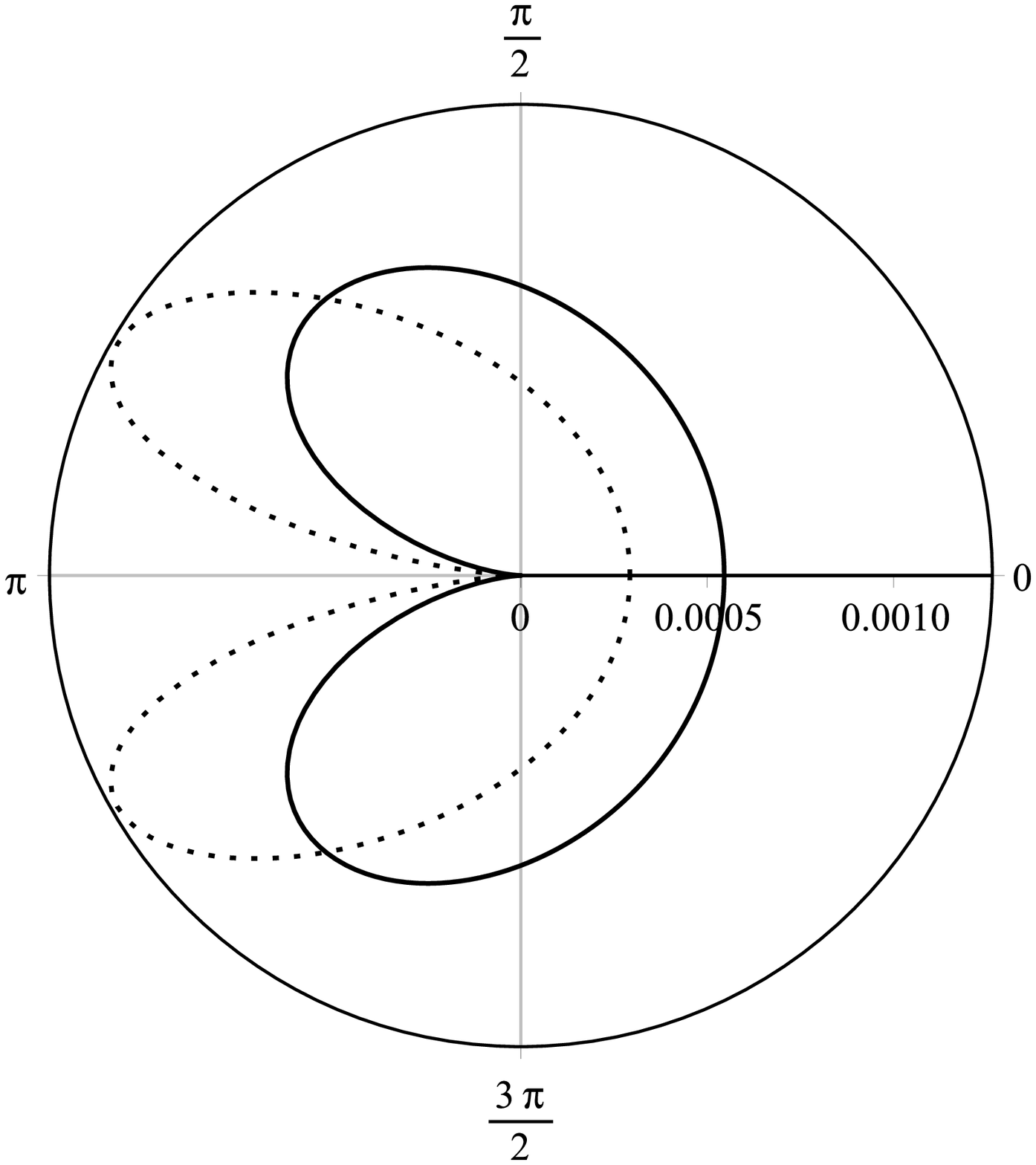}
\caption{Polar plot of the probability as a function of angle $\beta-\varphi$, in the case when helicity is not conserved for $k=0.1$. The solid line is for $p/p\,'=0.4$ and the point line for $p/p\,'=0.6$.}
\label{f11}
\end{figure}

Further we present the polar plots of the probability as a function of angle $\beta-\varphi$. In Figs.(\ref{f8})-(\ref{f9}), are given the polar plots of the probability in the case when the helicity is conserved. The probability is larger at angles very close to $\pi$ and when the momenta vectors are almost equal as moduli. The polar plots of the probability in the helicity nonconserving case are presented in Figs.(\ref{f10})-(\ref{f11}). Here the probability is maxim at small angles, except with the case when the momenta vectors are close as moduli, when the maximum is close to $\pi$ (see Fig.(\ref{f11})).

The polar plots give the possible directions in which the fermions could be emitted in the plane $(1,2)$, since the angles $\varphi\,,\beta$ are the polar coordinates of the momenta. The polar angles $\varphi\,,\beta$ are measured counterclockwise. If, for example, we take $\varphi=0$, then we can fix the orientation and direction for the momenta of the antiparticle $\vec{p}\,'$ . This means that $\vec{p}\,'$ points from the center of the circle toward the values $0$ on the trigonometric circle. In this case Figs.(\ref{f8}),(\ref{f9}) show the possible directions and orientations for the momenta of particle $\vec{p}$ and the possible values of $\beta$ around $\pi$ (the possible directions for $\vec{p}$ are in the ellipses in Figs.(\ref{f8}),(\ref{f9})). In the case of helicity conservation when the ratio $p/p\,'$ is close to unity, the vector $\vec{p}$ points from the center of the circle toward the value $\beta=\pi$ (see Fig. (\ref{f9})). When the ratio $p/p\,'$ becomes small, the momentum of the particle $\vec{p}$ could be oriented such that $\beta$ could take a large spectrum of values around $\pi$ (see Fig.(\ref{f8})).
\ \

\ \

\ \

\ \

\ \

The polar plots also show that, when helicity is conserved the momentum of the particle could not have the same orientation with the antiparticle momenta ($\varphi=\beta=0$), because in this case the probability is zero. Passing to the case when helicity is not conserved, we observe that taking $\varphi=0$ and fixing the direction and orientation of $\vec{p}\,'$, then the momenta of the particle $\vec{p}$ will be oriented such that the maximum of the probability is reached for $\beta=0$, in the case when the ratio $p/p\,'$ is small (see Fig.(\ref{f10})). When the ratio $p/p\,'$  approaches unity, the momenta of the particle could be oriented such that $\beta$ could increase up to values around $\pi$ as we can see from Fig.(\ref{f11}), with the observation that in $\beta=\pi$ the probability vanishes.

\section{Limit cases}
This section is dedicated to the study of the amplitudes/probabilities in the cases when the parameter $k=m/\omega$ takes large/small values. We start with the Minkowski limit that is obtained when the expansion parameter becomes zero or equivalently if we take $k=\infty$. The functions $f_{k}\left(\frac{p\,'}{p}\right)$ that define our amplitude (\ref{af}) vanish in this limit $\mathcal{A}_{e^-e^+}|_{k=\infty}=\mathcal{A}_{e^-e^+}(flat)=0$, since it depends on the factor $\cosh^{-1}(\pi k)$ which is highly convergent for large $k$. In the Minkowski limit the amplitude vanishes since the corresponding process from Minkowski QED is forbidden by the energy-momentum conservation laws. Because the probability is proportional with a factor $|\vec{p}+\vec{p\,'}|^{-4}$, it is clear that it will not be surprising if we recover the same Minkowski limit for the probability when the momenta $p\,,p\,'$ become very large. The graphical results also show that the probability is zero when $k\rightarrow \infty$.

The opposite case is obtained when $k=0$, this being the case when the expansion parameter approaches infinity (or the mass of the fermion is zero). Setting $k=0$ in the functions $f_{k}\left(\frac{p\,'}{p}\right)$ that define the amplitude and using Eq.(\ref{ff}), we finally obtain:
\begin{equation}
f_{0}\left(\frac{p\,'}{p}\right)=\frac{\pi}{1+\frac{p\,'}{p}}.
\end{equation}
From this simple equation we observe that this function becomes real  $f_{0}\left(\frac{p\,'}{p}\right)=f_{0}^*\left(\frac{p\,'}{p}\right)$ and that the amplitude is vanishing when the helicity conservation law is broken $(\lambda=\lambda\,')$. The amplitude equation in this limit is obtained if we observe that it is no longer necessary to use unit step functions, the final result being:
\begin{equation}\label{a0}
\mathcal{A}_{e^-e^+}|_{(k=0)}=
\frac{-ie}{2\pi^2|\vec{p}+\vec{p}\,\,'|^{2}}\,\frac{\delta_{\lambda,-\lambda}}{(p+p\,')}\,\,
\xi_{\lambda}^{+}(\vec{p}\,)[\vec{\sigma}\cdot(\vec{\mathcal{M}}\times(\vec{p}+\vec{p\,'}))]\eta_{-\lambda}(\vec{p\,'}).
\end{equation}
The fact that the amplitude is vanishing in the helicity nonconserving case for null mass fermions is also confirmed from our plots of the probability in terms of parameter $k$. In the null mass limit the only contribution in probability comes only from processes that preserve the helicity conservation law. The transition probability for null mass is obtained from the amplitude equation (\ref{a0}):
\begin{eqnarray}
\mathcal{P}_{e^-e^+}|_{(k=0)}&=&\sum_{\lambda}
\frac{e^2}{4\pi^4|\vec{p}+\vec{p}\,\,'|^{4}}\,\frac{1}{(p+p\,')^2}\,\,
|\xi_{\lambda}^{+}(\vec{p}\,)[\vec{\sigma}\cdot(\vec{\mathcal{M}}\times(\vec{p}+\vec{p\,'}))]\eta_{-\lambda}(\vec{p\,'})|^2\nonumber\\
&=&\frac{e^2\mathcal{M}^2}{4\pi^4|\vec{p}+\vec{p}\,\,'|^{4}}\,\frac{1}{(p+p\,')^2}(p-p\,')^2(1-\cos(\beta-\varphi)).
\end{eqnarray}

We obtain here that the probability in the zero mass limit vanishes if the particle and antiparticle have the same helicities, which corresponds to the nonconserving helicity in the process presented above. This result is expected, since the Dirac field for a null fermion mass is conformally invariant and the de Sitter metric is conformal with the Minkowski one.

Another relevant limit of our formulas is obtained when $k\gg1$, which corresponds to large mass comparatively with the expansion factor. This limit could be of interest since we can approximate the rate of pair production for weak gravitational fields. From (\ref{ff}), we obtain in the limit $k\gg1$:
\begin{equation}\label{km}
f_{k\gg 1}\left(\frac{p\,'}{p}\right)= \frac{2\pi \,e^{-\pi k}}{\left(1-\left(\frac{p\,'}{p}\right)^{2}\right)}\left(\left(\frac{p\,'}{p}\right)^{ik}-\left(\frac{p\,'}{p}\right)^{1-ik}\right).
\end{equation}
Using equation (\ref{km}) we can obtain the probability of pair production in the field of a magnetic dipole for $k\gg\,1$ (or $m \gg\omega$). The cases $p<p\,'$ and $p>p\,'$ are considered. These functions are very convergent for large $k$ and present an oscillatory behavior given by the factors $\left(\frac{p\,'}{p}\right)^{\pm ik}$. In this case the probability becomes:
\begin{eqnarray}
\mathcal{P}_{e^-e^+}&=&\frac{e^{2}\mathcal{M}^2e^{-2\pi k}}{4\pi^{4}}\frac{1}{(p^2-p\,'^2)^2(p^2+p\,'^2+2pp\,'\cos(\beta-\varphi))^2}\nonumber\\
&&\times\left[p^2g\left(\frac{p\,'}{p}\right)+
p\,'^{2}g\left(\frac{p}{p\,'}\right)\right]
\left\{
\begin{array}{cll}
(p-p\,')^2(1-\cos(\beta-\varphi))&{\rm for}&\lambda=-\lambda'\\
(p+p\,')^2(1+\cos(\beta-\varphi))&{\rm for}&\lambda=\lambda'
\end{array}\right.\label{ij}.
\end{eqnarray}
where the functions $g\left(\frac{p\,'}{p}\right)$ are defined below and $g\left(\frac{p}{p\,'}\right)$ is obtained when $p\leftrightarrows p\,'$:
\begin{eqnarray}
g\left(\frac{p\,'}{p}\right)=2+2\left(\frac{p\,'}{p}\right)^2-2\left(\frac{p\,'}{p}\right)^{1-2ik}-2\left(\frac{p\,'}{p}\right)^{1+2ik}
\pm\left(\frac{p\,'}{p}\right)^{2ik}\nonumber\\ \pm\left(\frac{p\,'}{p}\right)^{2+2ik}\pm\left(\frac{p\,'}{p}\right)^{-2ik}
\pm\left(\frac{p\,'}{p}\right)^{2-2ik}\mp4\left(\frac{p\,'}{p}\right)\left\{
\begin{array}{cll}
&{\rm for}&\lambda=-\lambda'\\
&{\rm for}&\lambda=\lambda'
\end{array}\right.
\end{eqnarray}
In this limit the probability is negligible, and our formulas suggest that for large $k$, the probability vanishes as $e^{-2\pi k}$.

A simple graphical analysis shows that the functions $g$ have an oscillatory behavior in terms of parameter $m/\omega$ that extends for large values of $m/\omega$. However, when replaced in the probability the exponent $e^{-2\pi k}$ will cancel the oscillatory behavior and the probability will become negligible for $m/\omega >>1$. This result shows that the probability dependence of parameter $m/\omega$ is mainly determined by the exponent factor $e^{-2\pi k}$. For this reason the terms from the probability formula that are responsible for the oscillatory behavior can be neglected, without altering the result, and finally we obtain:
\begin{eqnarray}\label{pmm}
&&\mathcal{P}_{e^-e^+}\simeq\frac{e^{2}\mathcal{M}^2e^{-2\pi k}}{\pi^{4}}\frac{1}{(p^2-p\,'^2)^2(p^2+p\,'^2+2pp\,'\cos(\beta-\varphi))^2}\nonumber\\
&&\times\left\{
\begin{array}{cll}
(p-p\,')^4(1-\cos(\beta-\varphi))&{\rm for}&\lambda=-\lambda'\\
(p+p\,')^4(1+\cos(\beta-\varphi))&{\rm for}&\lambda=\lambda'
\end{array}\right.\label{ij}.
\end{eqnarray}
This formula shows that the probability drops rapidly to zero as $m/\omega\sim1$ and this behavior is preserved for any given values of $\mathcal{M}$. The perturbative results presented here are valid only in the large expansion conditions from the early Universe as we can see from graphs (\ref{f1})-(\ref{f3}). In other words, the calculations are valid only when the gravitational fields are much stronger than the magnetic field. For this reason it will be interesting to study the situation when the magnetic fields are dominant or the case when the gravity and magnetism are both strong. This requires the use of a different metric than de Sitter and the best candidates are the metrics that describe the black holes. It is known that in the vicinity of a black hole, the magnetic fields are very strong. This situation could have dramatic consequences upon the pair production process, increasing significantly the rate of pair production. The first step in approaching this interesting subject will be the study of the Dirac equation with an external potential that generates a magnetic field in the Schwarzschild geometry. The results obtained so far for the free Dirac equation in Schwarzschild geometry \cite{ch,co} suggest that such a study could be addressed. The analysis presented in this paper represents the first step in studding the production of particles when the magnetic fields and the gravitational fields are present. If one considers strong magnetic fields, then the nonperturbative approach to the problem of pair production could be used. As far as we know, the nonperturbative approach to the problem of fermion production in the magnetic field in de Sitter geometry was not covered in the literature.

\section{Conclusions}
The process $vacuum \rightarrow e^-  + e^+$ in the presence of an external
magnetic field was studied using perturbative methods.
We found that the most probable transitions are those in which the fermion pair is
emitted perpendicular to the direction of the magnetic field. Also we found that the momentum modulus is no longer conserved in the process of pair production in the magnetic field in de Sitter geometry. The probability properties were studied
in the cases when helicity is conserved and when the helicity conservation law is broken. Our analysis led to the conclusion that
in the helicity conserving case the fermions could separate, while in the helicity nonconserving case it is more probable for the pair to annihilate into the vacuum.

Another result worth mentioning is that the pair production process is significant only in the strong
gravitational fields of the early Universe. Therefore even a small field of a magnetic dipole could produce fermions in conditions from the early Universe due to the large expansion of the space.
When the expansion factor vanishes $\omega = 0$, we can recover the Minkowski limit where this process is
forbidden by the energy-momentum conservation laws. Consequently our result cannot be extended to the present day conditions since the amplitude is vanishing in this case. It will be interesting to study the phenomenon of fermion production in magnetic fields in other geometries, since this subject received less attention in the literature and could have important implications in Astrophysics. Our analysis was done in the de Sitter metric, which is adequate for describing the conditions from the early Universe. We hope that the results presented in our paper concerning the particle production in magnetic fields will be of interest for cosmology and quantum fields in de Sitter space.

\section{Appendix}
Here we briefly present the main steps leading to the amplitude of pair production in magnetic field (\ref{af}).  Using the relation that connects Hankel functions and Bessel $K$ functions \cite{12,13,18}:
\begin{equation}\label{a2}
H^{(1,2)}_{\nu}(z)=\mp \left(\frac{2i}{\pi}\right)e^{\mp
i\pi\nu/2}K_{\nu}(\mp iz),
\end{equation}
we finally arrive at integrals of the type \cite{18}:
\begin{eqnarray}\label{a3}
&&\int_0^{\infty} dz
z^{-\lambda}K_{\mu}(az)K_{\nu}(bz)=\frac{2^{-2-\lambda}a^{-\nu+\lambda-1}b\,^{\nu}}{\Gamma(1-\lambda)}\Gamma\left(\frac{1-\lambda+\mu+\nu}{2}\right)\Gamma\left(\frac{1-\lambda-\mu+\nu}{2}\right)\nonumber\\
&&\times\Gamma\left(\frac{1-\lambda+\mu-\nu}{2}\right)\Gamma\left(\frac{1-\lambda-\mu-\nu}{2}\right)
\,_{2}F_{1}\left(\frac{1-\lambda+\mu+\nu}{2},\frac{1-\lambda-\mu+\nu}{2};1-\lambda;1-\frac{b^2}{a^2}\right),\nonumber\\
&&Re(a+b)>0\,,Re(\lambda)<1-|Re(\mu)|-|Re(\nu)|.
\end{eqnarray}
In our case $\lambda=-1$, and the second condition for convergence
is satisfied. We also observe that in our case $a,b$ are complex.
For solving our integrals we add to $a$ a small real part
$a\rightarrow a+\epsilon$, with $\epsilon>0$, and in the  end we
take the limit $\epsilon\rightarrow 0$. This assures the
convergence of our integral and will correctly define the unit
step functions and $f_{k}$ functions. No notable differences appear if we
evaluate the integrals directly with imaginary $a,b$.

The final form of the functions $f_{k}$ is obtained if we use the following identity between hypergeometric functions \cite{12,18}:
\begin{eqnarray}\label{hy}
_{2}F_{1}(a,b;c;z)&=& \frac{\Gamma(c)\Gamma(c-a-b)}{\Gamma(c-a)\Gamma(c-b)}\,_{2}F_{1}(a,b;a+b-c+1;1-z)\\ \nonumber
&&+(1-z)^{c-a-b}\,\frac{\Gamma(c)\Gamma(a+b-c)}{\Gamma(a)\Gamma(b)}\,_{2}F_{1}(c-a,c-b;c-a-b+1;1-z),
\end{eqnarray}
and the well known relations between gamma Euler functions \cite{12,18},
\begin{equation}
\Gamma(1+z)=z\Gamma(z),\,\,\Gamma(1-z)\Gamma(z)=\frac{\pi}{\sin(\pi z)}.
\end{equation}

\textbf{Acknowledgements}

Mihaela-Andreea B\u aloi was supported by the strategic grant POSDRU/159/1.5/S/137750, Project "Doctoral and Postdoctoral programs support for increased competitiveness in Exact Sciences research" cofinanced by European Social Fund within the Sectoral Operational Programme Human Resources Development 2007-2013.

Cosmin Crucean was supported by a grant of the Romanian National Authority for Scientific Research,
Programme for research-Space Technology and Advanced Research-STAR, project nr. 72/29.11.2013 between
Romanian Space Agency and West University of Timisoara.


\begin{thebibliography}{99}
\bibitem{0}
 C. W. Misner, K. S. Thorne and J. A. Wheleer, {\em Gravitation}
(W. H. Freeman and Company New York, 1973).
\bibitem{1}
J. D. Jackson ,{\em Classical Electrodynamics}, (John Wiley and Sons Ltd. 1962 ); W. Greiner, {\em Classical Electrodynamics}, (Springer 1998).
\bibitem{2}
A. H. Guth, {\em Phys. Rev. D} \textbf{23}, 347 (1981).
\bibitem{3}
L. F. Abbott and S. Y. Pi, {\em Inflationary Cosmology} (World Scientific, Singapore, 1986).
\bibitem{4}
M. S. Turner and L. M. Widrow, {\em Phys. Rev. D} \textbf{37}, 2743 (1988).
\bibitem{5}
G. B. Field and S. M. Carroll, {\em Phys. Rev. D} \textbf{62}, 103008 (2000).
\bibitem{6}
S. Kawati and A. Kokado, {\em Phys. Rev. D} \textbf{39}, 2959 (1989).
\bibitem{7}
S. Kawati and A. Kokado, {\em Phys. Rev. D} \textbf{39}, 3612 (1989).
\bibitem{8}
S. P. Gavrilov and D. M. Gitman, {\em Phys. Rev. D} \textbf{87}, 125025 (2013).
\bibitem{9}
H. K. Lee and Y. Yoon, {\em Mod. Phys. Lett. A} \textbf{22} (2007).
\bibitem{mo}
D. Marolf, I. A. Morrison, M. Srednicki, {\em Class. Quant. Grav.} \textbf{30}, 155023 (2013).
\bibitem{mo1}
D. Marolf and I. A. Morrison,  {\em Phys. Rev. D} \textbf{82} , 105032 (2010).
\bibitem{mo2}
D. Marolf and I. A. Morrison,  {\em Phys. Rev. D} \textbf{84} , 044040 (2011).
\bibitem{ak}
E. T. Akhmedov,  {\em Int. J. Mod. Phys. D} \textbf{23} ,  1430001 (2014).
\bibitem{10}
E. N. Parker, {\em Cosmological Magnetic Fields} (Clarendon, Oxford, England, 1979).
\bibitem{11}
Ya. B. Zel'dovich, A. A. Ruzmaikin and D. D. Sokoloff, {\em Magnetic fields in Astrophysics} (Gordon and Breach, New York, 1983).
\bibitem{12}
M. Abramowitz and I. A. Stegun, {\em Handbook of Mathematical Functions} (Dover, New York, 1964).
\bibitem{13}
G. N. Watson, {\em Theory of Bessel Functions} (Cambridge University Press, 1922).
\bibitem{14}
N. D. Birrel and P. C. W. Davies,  {\em Quantum Fields in Curved Space} (Cambridge University Press, Cambridge 1982).
\bibitem{fo}
N. D. Birrel, P. C. W. Davies and L. H. Ford, {\em J. Phys. A} \textbf{13}, 961 (1980).
\bibitem{15}
S. Weinberg, {\em The Quantum Theory of Fields}  (Cambridge University Press, Cambridge, 1995).
\bibitem{16}
S. Drell and J. D. Bjorken, {\em Relativistic Quantum Fields} (Mc Graw-Hill Book Co., New York 1965).
\bibitem{17}
L. Landau and E. M. Lifsit, {\em Theorie Quantique Relativiste} (Mir Moscou 1972).
\bibitem{18}
I. S. Gradshteyn and I. M. Ryzhik {\em Table of integrals, series and products} (Academic Press, 2007).
\bibitem{19}
Ion I. Cot\u{a}escu, {\em Phys. Rev. D} \textbf{65}, 084008 (2002).
\bibitem{20}
Crucean Cosmin, {\em Phys. Rev. D} \textbf{85}, 084036 (2012).
\bibitem{21}
Ion I. Cot\u{a}escu, C. Crucean, {\em Phys. Rev. D} \textbf{87}, 044016 (2013).
\bibitem{22}
Ion I. Cot\u{a}escu, C. Crucean, {\em Progress of Theor. Phys.} \textbf{124}, 1051 (2010).
\bibitem{23}
M. A. B\u{a}loi, {\em Mod. Phys. Lett. A} \textbf{29}, 1450138 (2014).
\bibitem{24}
E. Schr\" odinger, {\em Physica} \textbf{6}, 899 (1939).
\bibitem{25}
L. Parker, {\em Phys. Rev. Lett.} \textbf{21}, 562 (1968).
\bibitem{26}
L. Parker, {\em Phys. Rev.} \textbf{183}, 1057 (1969).
\bibitem{27}
L. Parker, {\em Phys. Rev. D} \textbf{3}, 346 (1971).
\bibitem{28}
K. H. Lotze, {\em Nucl. Phys. B} \textbf{312}, 673 (1989).
\bibitem{29}
K. H. Lotze, {\em Class. Quant. Grav.} \textbf{2}, 351 (1988).
\bibitem{30}
K. H. Lotze, {\em Class. Quant. Grav.} \textbf{5}, 595 (1985).
\bibitem{31}
V. M. Villalba, {\em Phys. Rev. D} \textbf{52}, 3742 (1995).
\bibitem{32}
J. Garriga, {\em Phys. Rev. D} \textbf{49}, 6343 (1994).
\bibitem{33}
I. L. Buchbinder, E. S. Fradkin and D. M. Gitman, {\em Fortschr. Phys.} \textbf{29}, 187 (1981).
\bibitem{34}
I. L. Buchbinder and L. I. Tsaregorodtsev, {\em Int. J. Mod. Phys. A} \textbf{7}, 2055 (1992).
\bibitem{35}
J. Haro and E. Elizalde, {\em J. Phys. A} \textbf{41}, 372003 (2008).
\bibitem{36}
P. Candelas and D. J . Raine, {\em Phys. Rev. D} \textbf{12}, 965 (1975).
\bibitem{37}
A. O. Barut and I. H. Duru, {\em Phys. Rev. D} \textbf{36}, 3705 (1987).
\bibitem{ch}
S. Chandrasekhar, {\em The Mathematical Theory of Black
Holes} (Oxford University Press, New York, 1983).
\bibitem{co}
I. I. Cot\u aescu, {\em Mod. Phys. Lett. A} {\bf 22}, 2493  (2007).

\end{thebibliography}
\end{document}